\documentclass[12pt, twoside]{article}
\usepackage{a4wide,amssymb,cite}

\usepackage{multicol,bbm}
\usepackage{amsmath,amssymb,setspace,subfigure,epsfig}
\usepackage{amsfonts}
\usepackage{graphicx}

\usepackage{color}

\newcommand{\be}{\begin{equation}}
\newcommand{\ee}{\end{equation}}
\newcommand{\bea}{\begin{eqnarray}}
\newcommand{\eea}{\end{eqnarray}}

\newcommand{\dis}[1]{\begin{equation}\begin{split}#1\end{split}\end{equation}}

\newcommand{\gev}{\,\textrm{GeV}}

\newcommand{\mgra}{{m_{3/2}}}
\newcommand{\mgauge}{{M_{1/2}}}
\newcommand{\tb}{{\tan\beta}}

\newcommand{\eq}[1]{Eq.~(\ref{#1})}

\newcommand{\fig}[1]{Fig.~\ref{#1}}
\newcommand{\oh}{{\Omega h^2}}

\newcommand{\tildeq}{\tilde{q}}
\newcommand{\mHu}{m_{{\tilde h}_u}}
\newcommand{\mHd}{m_{{\tilde h}_d}}

\begin{document}

\begin{titlepage}

\hfill  FTUAM 09/2, IFT-UAM/CSIC-09-05
\rightline{January 2009}

\begin{centering}
\vspace{1cm}
{\Large  {\bf $U(1)_R$-mediated supersymmetry breaking from a six-dimensional flux compactification}}\\

\vspace{1.5cm}

 {\bf Ki-Young Choi} $^{a,b,*}$ and  {\bf Hyun Min Lee} $^{c,**}$ \\
\vspace{.2in}

{\it $^{a}$Departamento de F\'{\i}sica Te\'{o}rica C-XI,
        Universidad Aut\'{o}noma de Madrid \\ 
        Cantoblanco, 28049 Madrid, Spain.\\
\vspace{3mm}        
$^{b}$Instituto de F\'{\i}sica Te\'{o}rica UAM/CSIC,
        Universidad Aut\'{o}noma de Madrid \\ 
        Cantoblanco, 28049 Madrid, Spain.\\
\vspace{3mm}
$^{c}$Department of Physics and Astronomy, McMaster University \\
Hamilton, Ontario L8S4M1, Canada. \\}

\end{centering}
\vspace{2cm}

\begin{abstract}
\noindent
We study the $U(1)_R$-mediated supersymmetry breaking in a flux compactification of 6D chiral gauged supergravity with codimension-two branes.
We consider a concrete model with manifest $U(1)_R$ invariance for moduli stabilization and visible sector in the context of 4D effective supergravity with gauged $U(1)_R$ and determine soft scalar masses in the visible sector mainly by a nonzero $U(1)_R$ D-term.
We obtain a low energy superparticle spectrum and discuss on the implications of the obtained non-universal scalar soft masses on the SUSY phenomenology such as dark matter relic abundances.

\end{abstract}

\vskip 1cm

\vskip 1cm

\vspace{2cm}
\begin{flushleft}

$^*~$E-mail: kiyoung.choi@uam.es \\
$^{**}~$E-mail: hminlee@mcmaster.ca

\end{flushleft}
\end{titlepage}

\section{Introduction}

Supersymmetry(SUSY)\cite{nilles} has been one of the promising candidates beyond the Standard Model(SM), in particular,
as a solution to the hierarchy problem in the SM Higgs sector.
Soft mass parameters that break SUSY while keeping the absence of quadratic divergences, however,
are subject to strong experimental constraints such as Flavor Changing Neutral Currents and CP violations, the so called SUSY flavor problem.
Therefore, to get problem-free soft mass parameters, one has to go beyond the simple 4D gravity mediation where the SUSY flavor problem is not explained.

Although Baryon/Lepton(B/L) number conservation is a result of gauge symmetry in the SM, it is not true of the 
the Minimal Supersymmetric Standard Model(MSSM) any more because the dimension-four and dimension-five B/L number violating operators are compatible with SM gauge symmetry. Even with $R$-parity conservation, the dimension-five operators are allowed. In this regard, a continuous $U(1)_R$ symmetry, that is a global symmetry of ${\cal N}=1$ SUSY algebra, can forbid both the dimension-four and dimension-five operators\cite{BL}. A continuous $U(1)_R$ symmetry may also solve the $\mu$ problem\cite{muproblem}.

However, any continuous symmetry should be gauged in order for quantum gravity effect like virtual black holes not to spoil the continuous symmetry. Thus, we consider the case where a continuous $U(1)_R$ symmetry appears 
as a local or gauge symmetry\footnote{We note that a continuous $U(1)_R$ symmetry can be an accidental global symmetry at lower orders as a result of discrete $R$ symmetry\cite{nillesratz}.}. Since components fields in a chiral multiplet have different $R$-charges, the local $U(1)_R$ symmetry can be realized only in the supergravity context\cite{u1rgauge}.
Since the MSSM fermions are charged under the $U(1)_R$, one has to take into account the anomaly cancellation conditions\cite{dreiner,castano}. When the $U(1)_R$ is gauged in 4D supergravity, there appears a nonzero Fayet-Iliopoulos term\cite{u1rgauge}.
A possibility of having the $U(1)_R$ as a new source for D-term SUSY breaking in the visible sector was also considered in 4D supergravity\cite{u1rmed}. 

In this paper, we consider a 4D effective supergravity with gauged $U(1)_R$, that is derived from a supersymmetric flux compactification with codimension-two branes in 6D chiral gauged supergravity\cite{leepapa,lee}. The background geometry preserves 4D ${\cal N}=1$ SUSY and it is featured by the unwarped product of 4D Minkowskian space and two extra dimensions that are {\it spontaneously} compactified on a football or rugby-ball due to a bulk $U(1)_R$ gauge flux\cite{background,lee}. Two codimension-two branes with nonzero tension are situated at the conical singularities of the internal dimensions, i.e. the poles of the football. 
From the 4D perspective, the bulk flux induces an additional FI term with $T$-modulus dependence that cancels the large constant FI term at the vacuum. Brane multiplets, both chiral superfields and vector superfields, were introduced on the codimension-two branes, being compatible with the bulk gauged supergravity\cite{lee}. The MSSM fields are assumed to be localized on the visible brane while the hidden sector fields are to be localized on the hidden brane.  
Dimensionally reducing to 4D on the supersymmetric football background, the 4D effective gauged supergravity with brane multiplets was also derived\cite{lee}. 

We first present a $U(1)_R$-anomaly free model with the MSSM fields where the SM-$U(1)_R$ anomalies are cancelled {\it $\grave{a}$ la} Green-Schwarz mechanism\cite{gs}.
By fixing the anomaly coefficients with the universal conditions on the brane-localized Green-Schwarz terms at the GUT scale, we find that it is possible to cancel the SM-$U(1)_R$ mixed anomalies without introducing additional SM non-singlets and 
there is a single family of solutions to the family-independent $R$-charges for the MSSM fields.
We don't deal with the explicit cancellation of pure $U(1)_R$ anomalies because it could be done independent of the cancellation of SM-$U(1)_R$ anomalies. This model was already discovered in Ref.~\cite{dreiner} but the $R$-charges for individual fields were not shown there because the authors in Ref.~\cite{dreiner} were mainly interested in the cancellation of pure $U(1)_R$ anomalies with a small number of SM singlets. 

In our 4D effective supergravity, the flux-induced $U(1)_R$ D-term only fixes one modulus, the $T$-modulus, leaving the other modulus, the $S$-modulus, unfixed.
Thus, we consider a concrete model for moduli stabilization by introducing a bulk gaugino condensate that generates an $S$-dependent effective superpotential\cite{lee}.
Since the $S$-modulus is neutral under the $U(1)_R$, the $U(1)_R$ invariance of the non-perturbative superpotential needs an inclusion of bulk matter fields that are charged under the bulk condensing gauge group.
In order to stabilize the matter fields, we couple to the matter fields a singlet chiral multiplet localized on the hidden brane in a $U(1)_R$-invariant fashion. 
However, for the $U(1)_R$-invariant superpotential with the global SUSY conditions, there always exists a flat direction. 
Therefore, one has to introduce a $U(1)_R$-breaking term in the superpotential in order to lift up the flat direction.
To this, we add a constant term in the superpotential which can be induced by a spontaneous breaking of the $U(1)_R$ symmetry in another sector without breaking the local SUSY. 
Focusing on the case that the supersymmetric masses for the $S$-modulus and the singlet scalars are larger than the gravitino mass, we show that it is possible to fix the $S$-modulus at order one and the singlet scalars at small VEVs approximately by using their SUSY conditions. On the other hand, as the superpotential is independent of the $T$-modulus, the F-term for the $T$-modulus does not vanish. 
After moduli stabilization, the vacuum energy becomes negative so
we need a hidden-brane F-term uplifting potential for a vanishing vacuum energy.
Finally, from the $T$-modulus minimization of the scalar potential, we show that the $U(1)_R$ D-term is nonzero. 
Thus, we find that the $U(1)_R$ mediation is a dominant source of SUSY breaking, generating soft masses of order the gravitino mass for visible scalars with nonzero $R$-charge\cite{lee}. 

Scalar soft mass squareds can be positive only for negative $R$-charges.
Then, being compatible with the consistent $R$-charges of the MSSM fields,
we find that there is a parameter space of $R$-charges that allows for all the squarks and sleptons to have positive soft mass squareds. In this case, the soft mass squareds for two Higgs doublets are negative. 
On the other hand, since the tree-level gauge kinetic functions of the brane-localized gauge fields are constant, the gaugino masses are zero at tree level. However, the $T$-dependent anomaly counterterms on the brane can induce universal gaugino masses at the GUT scale because of the nonzero $T$-modulus F-term for the $T$-independent superpotential. 
For a reasonably small $U(1)_R$ gauge coupling, the gaugino masses can be of order the gravitino mass.
For a phenomenological discussion on the $U(1)_R$ mediation at low energy, we take the gaugino mass to be a free parameter.

In the $U(1)_R$ mediation, there are five free parameters given at the GUT scale: $m_{3/2}$, $M_{1/2}$, $\tilde q$, ${\rm tan}\beta$ and ${\rm sign}(\mu)$ where $\tilde q$ is the $R$-charge of doublet squarks.
Consequently, we discuss about the impact of the obtained non-universal scalar soft masses on the SUSY phenomenology, in particular, the low-energy SUSY spectrum and the dark-matter constraints on the model parameters. The relic density of neutralino can match the WMAP bound in the pseudoscalar Higgs annihilation funnels. In this case, heavier neutral $H^0$ and $A$ Higgs bosons and charged $H^\pm$ Higgs bosons can be rather light so there should be an interesting experimental signature associated with the decays of heavier Higgs bosons produced at the LHC.  
Another characteristic feature of the $U(1)_R$ mediation is that in the stau-neutralino coannihilation region, gravitino is always a LSP and a candidate of dark matter while neutralino or stau is NLSP. 

The paper is organized as follows.
We first give a brief review on the 4D effective gauged supergravity derived from a flux compactification in 6D chiral gauged supergravity. Then we present a consistent set of $R$-charges of the MSSM fields for a $U(1)_R$-anomaly model.
We continue to discuss on the moduli stabilization and determine the soft masses in the visible brane at the minimum of the moduli scalar potential. In next section, we give a detailed discussion on the $U(1)_R$ SUSY phenomenology, 
focusing on the dark-matter constraints. Finally, a conclusion is drawn.
There are two appendices: one deals with the K\"ahler metric and the F-terms while
the other provides the general expressions for the scalar potential and the soft masses in 4D effective gauged supergravity.

\section{The gauged $U(1)_R$ supergravity}

We consider a flux compactification in 6D chiral gauged supergravity\cite{NS} 
where two extra dimensions are compactified on a supersymmetric football\cite{leepapa}.
The bulk fields in 6D chiral gauged supergravity are composed of the minimal gravity multiplet and an abelian vector multiplet.
The minimal gravity multiplet is a gravity multiplet($e^A_M,\psi_M,B^+_{MN}$) and a tensor multiplet($\phi,\chi,B^-_{MN}$),
and the vector multiplet($A_M,\lambda$) is needed to gauge the $U(1)_R$ symmetry.
There are two codimension-two branes with nonzero equal tensions located at the poles of the football. 
Being consistent with the bulk SUSY\cite{lee}, we introduce chiral multiplets $Q_i$ and vector multiplets on the visible brane
and chiral multiplets $Q',\varphi$ on the hidden brane. 
We also consider an SM neutral chiral multiplet $X$ coming from the bulk. 

For the flux compactification on a football with brane matters, 
the K\"ahler potential in 4D effective supergravity is identified\cite{lee} with 
\bea
K&=&-\ln\Big(\frac{1}{2}(S+S^\dagger)\Big) -\frac{2\xi_R}{M^2_P}V_R\nonumber \\
&&-\ln\Big(\frac{1}{2}(T+T^\dagger-\delta_{GS}V_R)-Q^\dagger_i e^{-2r_ig_R V_R}Q_i-Q^{'\dagger}e^{-2r'g_RV_R}Q'-\varphi^\dagger e^{-2r_{\varphi} g_R V_R}\varphi\Big) \nonumber \\
&&+X^\dagger e^{-2r_{X}g_R V_R}X   \label{kahlergen}
\eea
where we took the minimal K\"ahler potential for the bulk chiral superfield $X$.
Here the Green-Schwarz parameter is $\delta_{GS}=8g_R$ and the constant FI term is parametrized by $\xi_R=\frac{1}{4}\delta_{GS}M^2_P$. Furthermore, $V_R$ is the $U(1)_R$ vector superfield and $r_I$ are the $R$ charges of the superfields $\Phi^I=(Q_i,Q',\varphi,X)$. The $U(1)_R$ gauge boson mass squared is given by $M^2_R=8g^2_RM^2_P$ via a Green-Schwarz mechanism.
In the above, the scalar components of the moduli supermultiplets are
\be
S=s+i\sigma, \quad T=t+|Q_i|^2+|Q'|^2+|\varphi|^2+ib.
\ee
Here the scalar components $s$ and $t$ are written as the mixture of the dilaton and the volume modulus as $s=e^{\psi+\frac{1}{2}\phi}$ and $t=e^{\psi-\frac{1}{2}\phi}$ where $\phi$ is the dilaton and $\psi$ is the volume modulus.
Moreover, the axial scalar components, $\sigma$ and $b$, are derived from 
the relations, $e^f G_{\mu\nu\rho}=\epsilon_{\mu\nu\rho\tau}\partial^\tau\sigma$ and $b=-\frac{1}{2}\epsilon^{mn}{\cal B}_{mn}$, respectively, where $G_{\mu\nu\rho}$ is the field strength of the Kalb-Ramond(KR) field and ${\cal B}=B-\frac{1}{2}\langle A\rangle \wedge {\cal A}$ with $B$ being the KR field and $\langle A\rangle({\cal A})$ being the background VEV(fluctuation) of the $U(1)_R$ gauge boson. 
The brane chiral multiplets $Q_i$ can be also charged under the brane vector multiplets so we assume that all the MSSM fields are localized on the same codimension-two brane. 
The tree-level gauge kinetic functions for the bulk and brane vector multiplets are identified as $f_R=S$ and $f_W=1$, respectively. Consequently, the brane vector multiplets have no tree-level coupling to the bulk moduli while the brane chiral multiplet in the K\"ahler potential has a direct coupling to the $T$ modulus.
However, as will be shown later, the anomaly corrections to the brane gauge kinetic term have the $T$-modulus dependence.

The superpotential is composed of brane and bulk contributions as follows,
\be
W= W_1(Q_i) + W_2(Q',\varphi) +W_{\rm bulk}(S,T,X)+ W_{\rm mix}(\varphi,X).
\ee
The brane superpotentials $W_1, W_2$ do not depend on the moduli\cite{lee} and there is no tree-level coupling between 
the visible and hidden sectors because they are separated from each other geometrically in extra dimensions.
On the other hand, the bulk superpotential $W_{\rm bulk}$ can have the moduli dependence due to the bulk non-perturbative dynamics as will be discussed in the later section. Moreover, $W_{\rm mix}$ contains the couplings between the hidden sector and the bulk sector. For instance, it will be introduced for the stabilization of bulk scalar fields appearing in the gaugino condensates. However, we assume that there are no renormalizable couplings between $X$ and $Q_i$.

The brane chiral multiplet $Q_i$ having an $R$ charge $r_i$ transforms under the $U(1)_R$ 
with parameter $\Lambda$ (where ${\rm Re}\,\Lambda|_{\theta={\bar\theta}=0}=\Lambda_R$) as 
\be
Q_i\rightarrow e^{ir_ig_R\Lambda} Q_i
\ee
while the $U(1)_R$ vector multiplet transforms as 
\be
V_R\rightarrow V_R+\frac{i}{2}(\Lambda-\Lambda^\dagger). \label{gaugetransf}
\ee
The other chiral superfields $Q',\varphi,X$ transform similarly.
Gauge invariance of the $T$-dependent piece of the K\"ahler potential (\ref{kahlergen}) 
requires that, under the $U(1)_R$ gauge transformation, 
the $T$ modulus transforms nonlinearly as
\be
T\rightarrow T+\frac{i}{2}\delta_{GS} \Lambda.
\ee 
This results in a shift of the axion field, $b\rightarrow b+\frac{1}{2}\delta_{GS} \Lambda_R=b+4g_R \Lambda_R$.
Moreover, the effective superpotential taking an $R$ charge $+2$ transforms under the $U(1)_R$ as follows,
\be
W\rightarrow e^{2ig_R\Lambda} W.
\ee

\section{The $U(1)_R$ anomaly-free model}

When the $R$ charge of a scalar\footnote{We also name the $R$-charge of a chiral superfield by the one of a scalar partner.} is $r_i$, the $R$ charge of a fermionic superpartner differs by one unit as $r_i-1$.
The $R$ charge of each fermion is denoted by the corresponding name in the SM, for instance, $l$ for lepton doublets and 
$q$ for quark doublets, etc. Then, the $R$ charges of the sfermions are ${\tilde l}=l+1$ for slepton doublets 
and ${\tilde q}=q+1$ for squark doublets, etc. Here we assume that the $R$ charges are family-independent.

The bulk gravitino and the bulk $U(1)_R$ gaugino as well as the brane SM gauginos are charged under the $U(1)_R$.
Moreover, for generic $R$-charge assignments for chiral superfields, the matter fermions can be also charged. 
Therefore, in order for the $U(1)_R$ invariance to be guaranteed at the quantum level, 
the anomaly cancellation conditions must be satisfied.
In this section, we pursue the constraints coming from the anomaly cancellation and present a $U(1)_R$ anomaly-free model of the MSSM field contents with the help of a Green-Schwarz mechanism\cite{gs}.

\subsection{The anomaly conditions for the $U(1)_R$}

When the renormalizable Yukawa couplings respect the $U(1)_R$ symmetry, 
we need to satisfy the following conditions for the $R$ charges,
\bea
l+e+h_d&=&-1, \label{yukawalep}\\
q+d+h_d&=&-1,  \label{yukawadq}\\
q+u+h_u&=&-1.\label{yukawauq}
\eea

The $U(1)_R$ anomaly coefficients involving the SM gauge group are 
\bea
C_1&=&3\Big(\frac{1}{2}l+e+\frac{1}{6}q+\frac{4}{3}u+\frac{1}{3}d\Big)+\frac{1}{2}(h_d+h_u), \label{u1anom}\\
C'_1&=&3(-l^2+e^2+q^2-2u^2+d^2)-h^2_d+ h^2_u, \label{u1mixed}\\
C_2&=&3\Big(\frac{1}{2}l+\frac{3}{2}q\Big)+\frac{1}{2}(h_d+ h_u)+2, \label{su2anom}\\
C_3&=&3\Big(q+\frac{1}{2}u+\frac{1}{2}d\Big)+3.\label{su3anom}
\eea
The coefficients correspond to ${\rm tr}(R Y^2)$, ${\rm tr}(R^2Y)$, ${\rm tr}(RT^2_{SU(2)})$ and ${\rm tr}(RT^2_{SU(3)})$,
in order. 
On the other hand, the pure $U(1)_R$ anomalies, i.e. $U(1)_R^3$ anomalies and $U(1)_R$-gravity mixed anomalies, are, respectively, 
\bea
C_R&=&3(2l^3+e^3+6q^3+3u^3+3d^3)+2h^3_d+2h^3_u+16+\sum_m z^3_m, \label{ur3}\\
C'_R&=&3(2l+e+6q+3u+3d)+2(h_d+ h_u)-8+\sum_m z_m \label{urgg}
\eea
where $z_m$ are the $R$ charges of SM-singlet fermions. 

It has been shown that when $C_1=C'_1=C_2=C_3=0$, there is no solution of the consistent $R$ charges \cite{dreiner}. 
When there are additional SM non-singlets\cite{dreiner}, it is possible to have the anomalies cancelled.
On the other hand, when the renormalizable Yukawa couplings for some light generations are absent, the anomaly conditions can be solved but the $R$-charges turn out to be family-dependent\cite{dreiner}.

We focus on the case where the renormalizable Yukawa couplings are allowed and the SM anomalies are cancelled by a Green-Schwarz mechanism\cite{gs}. Even in this case, we need to have $C'_1=0$ because the $(U(1)_R)^2-U(1)_Y$ anomaly cannot be cancelled by the Green-Schwarz mechanism. 
Although one can show that the SM anomalies are cancelled by a Green-Schwarz mechanism\cite{dreiner}, it is nontrivial to check the cancellation of the pure $U(1)_R$ anomalies explicitly with a small number of the $R$-charged SM singlets.
So, in this paper, we don't deal with the pure $U(1)_R$ anomalies, just assuming that they are cancelled by multiple SM neutral
fermions in the hidden sector, independent of the anomalies involving the SM gauge group.

\subsection{The Green-Schwarz mechanism}

The anomalies coming from the fermions with nonzero $R$ charge is represented as the nonvanishing 
$U(1)_R$ gauge transform of the Lagrangian as
\be
\delta{\cal L}=\Lambda_R(x)\sum_{a=1}^3\frac{C_a}{8\pi^2}{\rm tr}(F_a {\tilde F}_a)
\ee
where $\Lambda_R$ is related to $\Lambda$ in eq.~(\ref{gaugetransf}) by ${\rm Re}\Lambda|_{\theta={\bar\theta}=0}=\Lambda_R$.
Then, in order for the $U(1)_R$ to be anomaly free, 
the Lagrangian must be supplemented with a brane-localized Green-Schwarz(GS) term, the variation of which is given as follows, 
\be
\delta{\cal L}_{GS}=-\Lambda_R(x)\frac{\delta_{GS}}{2}\sum_{a=1}^3k_a\frac{1}{2}{\rm tr}(F_a {\tilde F}_a) \label{GSterm}
\ee
where $k_a$ are the Kac-Moody levels of the gauge algebra and they are related to the anomaly coefficients as follows,
\be
\frac{C_a}{k_a}=2\pi^2\delta_{GS}. \label{universalrel}
\ee
In most string models constructed at level $k=1$ for non-abelian groups, $k_2=k_3=1$. 
In our case, however, we assume the higher level string models\cite{higherlevel} satisfying $k_2=k_3\neq 1$. 
Then we impose 
\be
C_2=C_3.  \label{anomcoeff1} 
\ee
In the presence of the GS term, the gauge kinetic functions of the brane vector multiplets are modified to
\bea
f_a=\frac{1}{g^2_{a,0}}+k_a T
\eea
where $g_{a,0}$ are the tree-level gauge couplings that are moduli-independent.
Consequently, at the energy scale of the $T$ modulus stabilization which is around the GUT scale, 
the gauge couplings read 
\be
\frac{1}{g^2_a}=\frac{1}{g^2_{a,0}}+k_a {\rm Re}T.\label{effgaugecoupling}
\ee
Thus, for unified tree-level gauge couplings with $g^2_{3,0}=g^2_{2,0}$ and $g^2_{1,0}=\frac{3}{5}g^2_{2,0}$ at the GUT scale, the favorable choice of $\sin^2\theta_W=\frac{3}{8}$ at the GUT scale requires $k_1=\frac{5}{3}k_2$ or the following via eq.~(\ref{universalrel}),
\be
C_1=\frac{5}{3}C_2.  \label{anomcoeff2}
\ee

\subsection{The anomaly-free model via the Green-Schwarz mechanism}

In this section, we show a $U(1)_R$ anomaly-free model with renormalizable Yukawa couplings and family-independent 
$R$-charges, with the help of the Green-Schwarz mechanism.

From the quark Yukawa couplings, (\ref{yukawadq}) and (\ref{yukawauq}),
we obtain
\be
q+\frac{1}{2}u+\frac{1}{2}d+1+\frac{1}{2}(h_d+  h_u)=0.
\ee
So, compared to eq.~(\ref{su3anom}), we get the relation between Higgsino $R$-charges as
\be
h_d+ h_u=-\frac{2}{3}C_3.\label{higgsrcharge}
\ee
Using the addition of eqs.~(\ref{u1anom}) and (\ref{su2anom}), and from (\ref{yukawalep}) and (\ref{yukawauq}),
we obtain 
\be
q+\frac{1}{2}u+\frac{1}{2}d-(h_d+ h_u+2)-\frac{1}{2}(C_1+C_2)=0.
\ee
Then, from eqs.~(\ref{su3anom}) and (\ref{higgsrcharge}), we find the relation between the anomaly coefficients as
\be
C_3=3+\frac{1}{2}(C_1+C_2).\label{anomrel}
\ee
Therefore, from the conditions, (\ref{anomcoeff1}) and (\ref{anomcoeff2}), the anomaly coefficients are
\be
C_1=-15, \quad\quad C_2=C_3=-9. \label {anomfix}
\ee
These conditions for the anomaly coefficients were also considered in Ref.~\cite{dreiner}.
After eqs.~(\ref{higgsrcharge}) and (\ref{anomrel}) are derived, there are five remaining conditions for the $R$-charges
for six parameters $(l,e,u,d,q,h)$: three Yukawa couplings and $C'_1=0$ and $C_2$ anomaly equation. 
Thus, we find that there is one parameter family of solutions to the $R$-charges:
\bea
l&=&-3q-\frac{28}{3}, \quad e=-\frac{3}{7}q -\frac83, \quad u=\frac{17}{7}q+4,  \nonumber \\
d&=&-\frac{31}{7}q-12, \quad h_d=\frac{24}{7}q+11, \quad  h_u=-\frac{24}{7}q-5. \label{fermionrcharge}
\eea
So, the $R$-charges of the scalar superpartners\footnote{We note that the tilded letters are used for all scalars including Higgs scalars.} are
\bea
\tilde{l}&=&-3\tilde{q}-\frac{16}{3}, \quad \tilde{e}=-\frac{3}{7}\tilde{q} - \frac{26}{21}, \quad \tilde{u}=\frac{17}{7}\tilde{q}+\frac{18}{7},  \nonumber \\
\tilde{d}&=&-\frac{31}{7}\tilde{q}-\frac{46}{7}, \quad \tilde{h}_d= \frac{24}{7}\tilde{q}+\frac{60}{7}, \quad \tilde{h}_u=-\frac{24}{7}\tilde{q}-\frac{4}{7}. \label{scalarrcharge}
\eea
Here, we note that, since ${\tilde h}_d+{\tilde h}_u=8$, the tree-level $\mu$ term is not allowed. 
Thus, the $\mu$ term must be generated by the VEV of a singlet $N$ with the superpotential coupling $W=\lambda_N N^k H_u H_d$
where the $R$-charge of the singlet given by $r_N=-\frac{6}{k}$ is negative for a positive $k$.
We note that it is also possible to generate the $\mu$ term from the K\"ahler potential\cite{muproblem}
with $K=N^{\frac{4}{3}k}H_u H_d+{\rm h.c.}$ but in this case we would get a suppressed $\mu$ term as $\mu\sim \langle N\rangle^{\frac{4}{3}k}m_{3/2}$.

For the $R$-charges for fermions obtained in (\ref{fermionrcharge}), from eqs.~(\ref{ur3}) and (\ref{urgg}),
the pure $U(1)_R$ anomalies are
\be
C_R=-\frac{157348}{9}-\frac{132480}{7}q-\frac{333720}{49}q^2-\frac{273375}{343}q^3+\sum_m z^3_m, 
\ee
\be
C'_R=-132-\frac{135}{7}q+\sum_m z_m.
\ee
The anomalies of zero modes of gravitino and $U(1)_R$ gaugino and zero modes of other $R$-charged bulk fermions should be cancelled by the flux-induced
4D anomaly terms coming from a bulk Green-Schwarz term\cite{lee}. For instance, the anomaly contributions of zero-mode gravitino
and $U(1)_R$ gaugino amount to $3+1=4$ in $C_R$ and $-21+1=-20$ in $C'_R$.
Therefore, after subtracting the bulk zero-mode contributions, the anomalies would come only from the MSSM fermions and SM-singlet fermions localized on the branes, so the anomaly cancellation conditions are 
\be
-\frac{157384}{9}-\frac{132480}{7}q-\frac{333720}{49}q^2-\frac{273375}{343}q^3+\sum_{m'} z^3_{m'}=0, 
\ee
\be
-112-\frac{135}{7}q+\sum_{m'} z_{m'}=0
\ee
where $m'$ denotes the SM-singlet brane fermions.
Inequivalently, in terms of the $R$-charge of squark doublets, $\tilde q$, and the $R$-charges of the SM-singlet brane sfermions, ${\tilde z}_{m'}$,
we rewrite the above conditions as
\be
-\frac{14123017}{3087}-\frac{2639565}{343}\tilde {q}-\frac{1515915}{343}{\tilde q}^2-\frac{273375}{343}{\tilde q}^3
+\sum_{m'} ({\tilde z}_{m'}-1)^3=0,
\ee
\be
-\frac{649}{7}-\frac{135}{7}{\tilde q}+\sum_{m'} ({\tilde z}_{m'}-1)=0.
\ee

\section{Moduli stabilization and soft masses}

Although the bulk flux makes some of moduli fixed, there remains a modulus that is not fixed yet.
In this section, we discuss the modulus stabilization with a bulk non-perturbative effect in 4D effective supergravity
and find that the interplay of the heavy $T$ modulus with the light modulus is crucial in determining the soft masses
as the light $S$ modulus does not couple to the visible sector in the tree level K\"ahler potential.

\subsection{Bulk gaugino condensates}

When there is neither non-perturbative bulk dynamics or brane-localized superpotential, 
the 4D scalar potential is obtained\cite{quevedo6dsusy,lee} as follows,
\bea
V_0&=&\frac{2g^2_RM^4_P}{s}\bigg(1-\frac{1}{t}\bigg)^2.\label{dtermonly}
\eea
Therefore, the $T$ modulus is stabilized at $t=1$ by the bulk $U(1)_R$ flux, i.e. the $U(1)_R$ D-term in 4D effective theory.
However, the $S$ modulus remains a flat direction so one needs a stabilization mechanism by some bulk non-perturbative dynamics.

Suppose that there is a gaugino condensate preserving the $U(1)_R$ invariance.
Then, including the SUSY breaking represented by $Q'$ localized on the hidden brane, 
we consider the effective superpotential\footnote{Compare to Ref.~\cite{lee} where double gaugino condensates without a constant superpotential were introduced in the $U(1)_R$ non-invariant form.} as
\be
W= f Q' + W_0 +W_{\rm dyn} \label{superpgen}
\ee
with
\be
W_{\rm dyn}=\frac{\lambda}{X^{n}}e^{-b S}+\lambda'\varphi^{p} X^2+\kappa \varphi^{q} \label{superpdyn}
\ee
where $X$ is a bulk chiral superfield with $R$-charge $r_{X}=-\frac{2}{n}$,
$\varphi$ is a brane chiral superfield with $R$-charge $r_{\varphi}=\frac{2(n+2)}{p n}=\frac{2}{q}$,
and $f,W_0,\lambda,b,\lambda'$ and $\kappa$ are constant parameters.

The more details on the parameters of the superpotential are in order.
First, $W_0$ is assumed to be given by the VEV of a superpotential term for SM-neutral chiral multiplets in another sector. When the $U(1)_R$ symmetry is broken spontaneously to give a nonzero $W_0$, the global SUSY conditions for the SM-neutral chiral multiplets are not satisfied
because of the consistency condition for the $U(1)_R$-invariant superpotential\cite{nelsonseiberg,nillesratz}.
Instead we consider the case where the local SUSY conditions are fulfilled.
For instance, suppose that a superpotential in another sector is given by $W_0=Y^{2/r_Y}{\hat W}(Z)$ where $Y$ is a bulk singlet chiral superfield with $R$-charge $r_Y$ and ${\hat W}(Z)$ is an arbitrary holomorphic function of a brane-localized or bulk singlet chiral superfield $Z$ with zero $R$-charge.
Then, the SUSY condition for $Z$ would stabilize the $Z$ scalar VEV giving a nonzero $\langle {\hat W}\rangle$
while the local SUSY condition for $Y$, $D_Y W_0=0$, determines the $Y$ scalar VEV as $|Y|^2=\frac{2}{|r_Y|}$ for $r_Y<0$. 
As will be shown later, for a nonzero $W_0$, the $U(1)_R$ D-term gives rise to a soft squared mass for the $Y$ scalar proportional to $-r_Y$. Therefore, for $|r_Y|\gg 1$, the soft squared mass for $Y$ can be positive and much larger than 
the gravitino mass, overcoming the instability of the local SUSY vacuum with a negative supersymmetric squared mass of order the gravitino mass\cite{kobayashi}. On the other hand, the axionic part of the $Y$ scalar is not determined by the local SUSY condition. One of linear combinations of the axionic part of the $T$ modulus and the one of $Y$ is absorbed by the $U(1)_R$ gauge boson while the other combination remains a flat direction. If $Y$ also transforms under a global $U(1)$ symmetry, the anomaly coupling of the axionic part of the $Y$ scalar to hidden gauge group would generate a potential for the remaining axion after integrating out the hidden gauge fields. In this case, a small violation of the global symmetry in ${\hat W}(Z)$ would be needed to stabilize the $Z$ scalars by the SUSY conditions.
Here we assume that the contribution of the $Y$ scalar VEV to the $U(1)_R$ D-term is cancelled by a different scalar VEV with opposite $R$-charge. In the following discussion, we just parametrize the $U(1)_R$ symmetry breaking by $W_0$ without considering an explicit model for that.

The first term of $W_{\rm dyn}$ stems from a bulk gaugino 
condensate\cite{ads,gauginocondense} containing the meson field $X$. 
Since the $S$-modulus is neutral under the $U(1)_R$, it is necessary to include the meson field with a nonzero $R$-charge in the gaugino condensate.
The last two terms of $W_{\rm dyn}$ come from the interactions with $\varphi$ localized at the hidden brane. The different form of the interaction term would not change the conclusion drawn in the next section as long as $X$ and $\varphi$ scalars are stabilized at small values.
Regarding the bulk gaugino condensate, in 4D effective $SU(N)$ SUSY QCD with $F$ flavors in the fundamental and antifundamental representations of $SU(N)$ where $F< N$, the parameters in the effective superpotential are related to the fundamental parameters as
$\lambda=(N-F)(M_*/M_P)^{(3N-F)/(N-F)}$ where $M_*$ is the unification scale, $n=\frac{2F}{N-F}$ and $b=\frac{8\pi^2}{N-F}$.
Finally, the hidden brane SUSY breaking parametrized by $f$ is needed to lift up to zero the negative vacuum energy generated after moduli stabilization as will be shown later.

\subsection{The effective scalar potential}

When SUSY is unbroken, $Q'$ is a flat direction. However, when SUSY is broken, 
the coupling of $Q'$ to other massive chiral superfields stabilizes $Q'$ at zero by radiative corrections\cite{seiberg}. 
We assume that the VEVs of $Q_i$ and $Q'$ vanish and $F^{Q_i}=0$ while the VEVs of $X$ and $\varphi$ are nonzero. 
Then, the scalar potential is given by
\be
V_0=V_F + V_D \label{scalarpot}
\ee 
where $V_F$ is the F-term potential obtained from eq.~(\ref{ftermpot1}) with the effective superpotential (\ref{superpgen}),
as follows,
\be
V_F=M^4_Pe^{|X|^2}\bigg(\frac{4s}{t}|{\hat F}_S|^2
+\frac{1}{s}|{\hat F}_{Q'}|^2+\frac{1}{s}|{\hat F}_{\varphi}|^2 
+\frac{1}{st}|{\hat F}_{X}|^2-\frac{2}{st}|W|^2\bigg) \label{explicitpot}
\ee
with $t=\frac{1}{2}(T+T^\dagger)-\varphi^\dagger \varphi$, and $V_D$ is the D-term potential\footnote{We omit a D-term on the hidden brane\cite{lee} because its realization is model-dependent.} 
obtained from eq.~(\ref{dtermpot}) as follows,
\be
V_D=\frac{1}{2}sD^2_R
\ee
with
\be
D_R=\frac{2g_R M^2_P}{s}\Big(1-\frac{1}{t}+\frac{1}{2t}r_{\varphi} |\varphi|^2+\frac{1}{2}r_{X} |X|^2\Big). \label{udtermsimple}
\ee
The hatted F-terms\footnote{See the genuine F-terms in Appendix B2 for comparison.} are
\bea
{\hat F}_S&=&\frac{\partial W}{\partial S}-\frac{1}{2s}W, \\
{\hat F}_{Q'}&=&\frac{\partial W}{\partial Q'}, \\
{\hat F}_{\varphi}&=&\frac{\partial W}{\partial \varphi}, \\
{\hat F}_{X}&=&\frac{\partial W}{\partial X}+X^\dagger W.
\eea
Here we note that since the superpotential is independent of the $T$-modulus, the $T$-modulus F-term contribution to the scalar potential is cancelled by a negative supergravity correction term as shown in Appendix B2 from eq.~(\ref{ftermpot}) to eq.~(\ref{ftermpot1}).
We also note that since ${\rm Im}T$ does not appear in the scalar potential, it is a massless scalar that is absorbed by the $U(1)_R$ gauge boson.

From eq.~(\ref{scalarpot}), the minimization conditions of the scalar potential with respect to the moduli and the scalar fields are
\bea
\frac{\partial V_0}{\partial T}&=&\frac{1}{2}M^4_P e^{|X|^2}\bigg[-\frac{4s}{t^2}|{\hat F}_S|^2-\frac{1}{st^2}|{\hat F}_{X}|^2+\frac{2}{st^2}|W|^2\bigg] \nonumber \\
&&+\frac{g_RM^2_P D_R}{t^2}\Big(1-\frac{1}{2}r_{\varphi}|\varphi|^2\Big), \label{mint}
\eea
\bea
\frac{\partial V_0}{\partial S}&=&M^4_P e^{|X|^2}\bigg[\frac{2}{t}|{\hat F}_S|^2-\frac{1}{2s^2}|{\hat F}_{Q'}|^2
-\frac{1}{2s^2}|{\hat F}_{\varphi}|^2-\frac{1}{2s^2t}|{\hat F}_X|^2-\frac{2}{st}W^\dagger {\hat F}_S \nonumber \\
&&+\frac{4s}{t}{\hat F}^\dagger_S \Big(\frac{\partial^2 W}{\partial S^2}-\frac{1}{2s}\frac{\partial W}{\partial S}\Big)+\frac{1}{st}{\hat F}^\dagger_{X} \Big(\frac{\partial^2 W}{\partial X\partial S}+X^\dagger \frac{\partial W}{\partial S}\Big)\bigg]-\frac{1}{4}D^2_R, \label{mins}
\eea
\bea
\frac{\partial V_0}{\partial X}&=&\Big(V_F+\frac{2M^4_P}{st}|W|^2 e^{|X|^2}+r_{X} g_R M^2_P D_R\Big)X^\dagger \nonumber \\
&&+M^4_P e^{|X|^2}\bigg[\frac{4s}{t}{\hat F}^\dagger_S \Big(\frac{\partial^2 W}{\partial X\partial S}-\frac{1}{2s}\frac{\partial W}{\partial X}\Big)+\frac{1}{s}{\hat F}^\dagger_\varphi \frac{\partial^2 W}{\partial X\partial \varphi} \nonumber \\
&&+ \frac{1}{st}{\hat F}^\dagger_{X} \Big(\frac{\partial^2 W}{\partial X^2}
+X^\dagger \frac{\partial W}{\partial X}\Big)-\frac{2}{st}W^\dagger {\hat F}_{X}\bigg], \label{minx}
\eea
\bea
\frac{\partial V_0}{\partial\varphi}&=&\bigg[M^4_P e^{|X|^2}\Big(\frac{4s}{t^2}|{\hat F}_S|^2+\frac{1}{st^2}|{\hat F}_{X}|^2-\frac{2}{st^2}|W|^2\Big)\nonumber \\
&&\quad+\frac{1}{t}r_{\varphi} g_R M^2_P D_R -\frac{2g_R M^2_P D_R}{t^2}\Big(1-\frac{1}{2}r_{\varphi}|\varphi|^2\Big)\bigg] \varphi^\dagger  \label{minvarphi}\\
&&+M^4_P e^{|X|^2}\bigg[-\frac{2}{t}{\hat F}^\dagger_S {\hat F}_{\varphi} +\frac{1}{s}{\hat F}^\dagger_{\varphi} \frac{\partial^2 W}{\partial\varphi^2} 
+\frac{1}{st}{\hat F}^\dagger_{X} \Big(\frac{\partial^2 W}{\partial \varphi \partial X}+X^\dagger \frac{\partial W}{\partial\varphi}\Big)-\frac{2}{st}W^\dagger {\hat F}_{\varphi}\bigg]. \nonumber
\eea

From the $T$-modulus minimization of the scalar potential (\ref{mint}), using the vanishing vacuum energy condition,
we determine the $U(1)_R$ D-term as
\be
D_R=-\frac{1}{2g_R M^2_P}\Big(V_F+M^4_P e^{|X|^2}\frac{|{\hat F}_{Q'}|^2}{s}+M^4_P e^{|X|^2}\frac{|{\hat F}_{\varphi}|^2}{s}\Big)
\Big(1+\frac{1}{2}r_{X}|X|^2\Big)^{-1}. \label{dtermdetermine}
\ee

\subsection{Moduli stabilization}

We first consider the stabilization of moduli for ${\hat F}_{Q'}=0$, i.e. $f=0$ in the full superpotential (\ref{superpgen}), and next discuss on the effect of ${\hat F}_{Q'}\neq 0$.
Let's see the minimization condition with respect to the $S$-modulus, (\ref{mins}).
The $U(1)_R$ D-term contribution in eq.~(\ref{mins}) is negligible from eq.~(\ref{dtermdetermine}) for a weak-scale gravitino mass. Thus, if ${\hat F}_X={\hat F}_\varphi=0$, the scalar potential is minimized with respect to the $S$ modulus approximately for ${\hat F}_S=0$.
When the supersymmetric masses of $X$ and $\varphi$ chiral multiplets are larger than their soft mass terms multiplied by $X^\dagger$ or $\varphi^\dagger$ in eqs.~(\ref{minx}) and (\ref{minvarphi}),
the other minimization conditions for scalars, eqs.~(\ref{minx}) and (\ref{minvarphi}), are also satisfied approximately
for ${\hat F}_X={\hat F}_\varphi={\hat F}_S=0$. 
The supersymmetric mass terms are $m_{X}\sim \Big|\frac{\partial^2 W}{\partial X^2}\Big|\sim \Big|\frac{W}{X^2}\Big|$ and $m_\varphi\sim \Big|\frac{\partial^2 W}{\partial \varphi^2}\Big|\sim \Big|\frac{W}{\varphi^2}\Big|$ and 
the mixing mass term is $|\frac{\partial^2 W}{\partial X\partial S}|\sim \Big|\frac{bW}{X}\Big|$, 
while their soft mass terms appearing in eqs.~(\ref{minx}) and (\ref{minvarphi}) are of order $|W|$ for the $U(1)_R$ D-term obtained from eq.~(\ref{dtermdetermine}). Therefore, for $|X|\ll 1$ and $|\varphi|\ll 1$, the supersymmetric masses of $X$ and $\varphi$ can be much larger than their soft masses.

We consider the stabilization of scalars, $X$ and $\varphi$, in more detail.
For small scalar VEVs, the scalar VEVs are stabilized dominantly by
the global SUSY conditions, $\frac{\partial W}{\partial X}=\frac{\partial W}{\partial \varphi}=0$. 
Thus, the global SUSY conditions give the scalar VEVs in terms of the condensation scale $\Lambda\equiv \frac{\lambda}{X^n}e^{-bS}$ as follows,
\bea
X^{-n}&= &c_X \Lambda,   \label{xvev} \\
\varphi^q&= & c_\varphi \Lambda  \label{varphivev}
\eea
with
\bea
c_\varphi&=&-\frac{np}{2\kappa q}, \\
c_X &=&\Big(\frac{2\lambda'}{n}\Big)^{\frac{n}{2}} c^{\frac{np}{2q}}_\varphi. \label{cx}
\eea
The condition (\ref{xvev}) does not determine the $X$ scalar VEV, rather fixing the $S$-modulus as
\be
{\rm Re}S= s=\frac{1}{b}\ln|c_X\lambda|, \quad \,\, {\rm Im}S=\frac{1}{b}(\theta-2m\pi) \label{smodulusmin}
\ee
with $e^{i\theta}\equiv\frac{c_X\lambda}{|c_X\lambda|}$ and $m$ being integer.
On the other hand, eq.~(\ref{varphivev}) gives a relation between the scalar VEVs.
Thus, the global SUSY conditions for the $U(1)_R$-invariant superpotential $W_{\rm dyn}$ leaves a flat direction.
For the global SUSY conditions for matter fields, one can show that
the superpotential containing the matter fields vanish at the vacuum, 
as $W_{\rm dyn}= (1+\frac{n}{2}-\frac{pn}{2q})\Lambda=0$
from the relation between assigned $R$-charges.
This shows the consistency condition for the $U(1)_R$-invariant superpotential $W_{\rm dyn}$ with vanishing global SUSY F-terms\cite{nillesratz}.

\begin{figure}[!t]
  \begin{center}
  \begin{tabular}{c c}
   \includegraphics[width=0.5\textwidth]{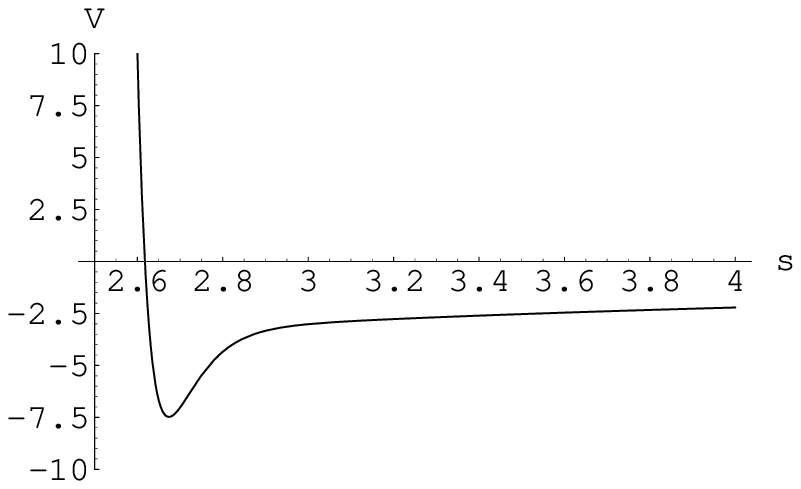}
&
   \includegraphics[width=0.5\textwidth]{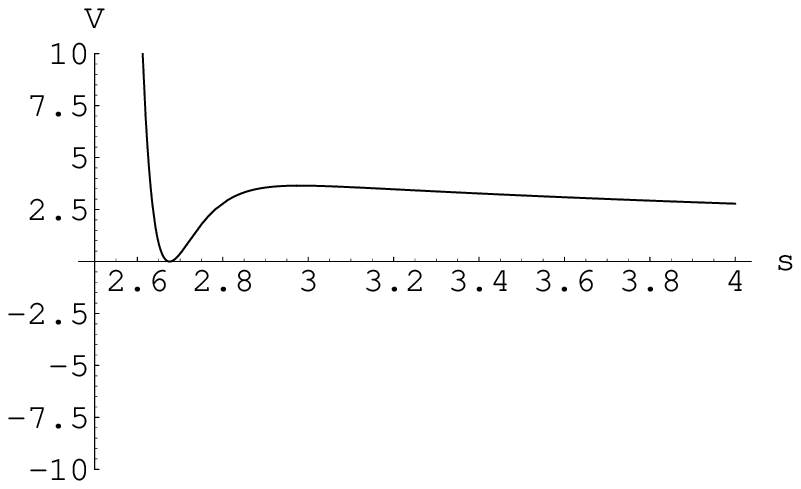}
   \end{tabular}
  \end{center}
 \caption{Plot of the scalar potential for $s={\rm Re}S$ with non-zero $W_0$ with $f=0$ (Left) 
and $f\ne 0$ (Right) to show the uplifting of the potential.
Here we used $ \lambda =0.01,\ b =15,\ 
\lambda^\prime = 10^{-7},\ p = 3,\ q= 1, n = 1,\ \kappa = -10^{-15}, 
W_0 = 10^{-16}$, with $f=0$ (Left) and $f= 1.413\times10^{-16}$ (Right).
The other scalar fields are fixed at the values given by SUSY vacuum 
in the text. They are approximately $t_0\simeq 1.00095, s_0\simeq 2.673, X_0\simeq -0.03087, \varphi_0\simeq -0.00187$.
The height of the scalar potential is multiplied by $10^{33}M^4_P$.
}
\label{plot_s_fzero}
\end{figure}

We now consider the $S$-modulus F-term. In the presence of a nonzero constant superpotential $W_0$, 
for ${\hat F}_X={\hat F}_\varphi=0$, a vanishing $S$-modulus F-term, ${\hat F}_S=0$, determines the condensation scale by the constant superpotential approximately as 
\be
\Lambda\simeq -\frac{W_0}{2bs} \label{condensescale}
\ee
where use is made of the global SUSY conditions, eqs.~(\ref{xvev}) and (\ref{varphivev}), 
for matter fields in computing the superpotential VEV. 
Thus, for $s\sim 1$ and $b\sim 10$, the condensation scale should be lower than the SUSY breaking scale $|W_0|$ by the order of magnitude. Consequently, for the fixed condensation scale in eq.~(\ref{condensescale}), from eqs.~(\ref{xvev}) and (\ref{varphivev}), we can fix the scalar VEVs too.
For $|X|\ll 1$ and $|\varphi|\ll 1$, from eqs.~(\ref{xvev}) and (\ref{varphivev}), we need the condition, $\frac{1}{|c_X|}\ll |\Lambda| \ll \frac{1}{|c_\varphi|}$, which corresponds to the following condition on the $\varphi$ couplings in the superpotential,
\be
\Big(\frac{n}{2|\lambda'|}\Big)^{\frac{n}{2}}\Big(\frac{2|\kappa|q}{np}\Big)^{\frac{np}{2q}}\ll |\Lambda| \ll \frac{2|\kappa| q}{np}.
\ee
Here we note that $\frac{np}{2q}=\frac{1}{2}(n+2)$ with $\frac{1}{2(F+1)}\leq \frac{1}{n+2}<\frac{1}{2}$ from the relation $n=\frac{2F}{N-F}$. Therefore, if $\frac{1}{n+2}$ is not so small, we need a hierarchy, 
$|\Lambda|\ll |\kappa|\ll |\lambda'|$.
For instance, we consider the case with a weak-scale gravitino for $|W_0|=\frac{m_{3/2}}{M_P}\sim 10^{-16}$. 
Then, from eq.~(\ref{condensescale}), $|\Lambda|\sim 10^{-17}$ for $bs\sim {\cal O}(10)$.
If we take $bs=37$ for $e^{-bs}=10^{-17}$, i.e. $|c_X\lambda|\sim 10^{17}$
from eq.~(\ref{smodulusmin}), from the definition of $\Lambda$, we get $|X^n|\sim |\lambda|=(N-F)(M_*/M_P)^{(3N-F)/(N-F)}\sim 0.01$
for $M_*/M_P\sim 10^2$ and $\frac{3N-F}{N-F}\sim 1$. From eq.~(\ref{varphivev}), $|\varphi^q|\sim |\Lambda/\kappa|\sim 0.01$ for $\kappa\sim 10^2|\Lambda|$. Therefore, from $|c_X|\sim 10^{17}/|\lambda|\sim 10^{19}$ with eq.~(\ref{cx}), we get $|\lambda'|\sim 10^{8/n}|\kappa|\sim 10^8|\kappa|$ for $n\sim 1$.
Consequently, in this example, we can get small scalar VEVs as $|X|\sim |\varphi|\sim 0.01$ for $n\sim q\sim 1$, and the needed hierarchy for the parameters in the superpotential is $|\kappa|/|\lambda'|\sim 10^{-8}|\Lambda|$ with $|\kappa|\sim 10^2|\Lambda|$.

\begin{figure}[!t]
  \begin{center}
  \begin{tabular}{c}
   \includegraphics[width=0.5\textwidth]{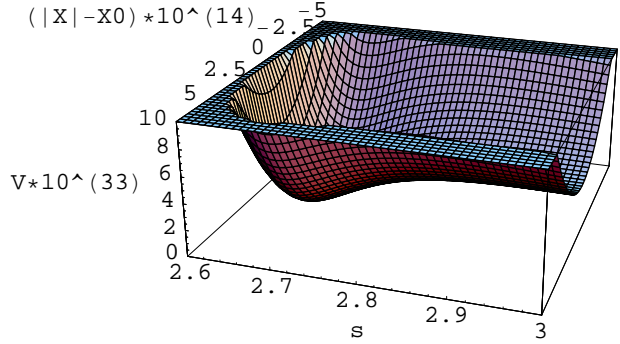}
   \end{tabular}
  \end{center}
 \caption{Plot of the scalar potential for $s=\textrm{Re} S$ and $|X|-X_0$, 
where $X_0$ is the VEV of $X$ with SUSY vacuum. Note the rescaled axes.
}
\label{plot_3d}
\end{figure}

Up to now, we have set ${\hat F}_{Q'}=0$ for the moduli of stabilization.
After stabilizing the moduli, however, from eq.~(\ref{scalarpot}), the vacuum energy becomes negative, so we must have ${\hat F}_{Q'}\neq 0$ to lift up the vacuum energy to zero. 
So, we now comment on the effect of a hidden brane SUSY breaking on the moduli stabilization.
The hidden brane F-term leads to an additional potential for the $S$-modulus. However, the minimum determined by the ${\hat F}_S=0$ condition can be shifted a little bit by the hidden brane F-term the scale of which is set by the gravitino mass. 
The reason is the following.
The supersymmetric mass of the $S$-modulus is given by $\Big|\frac{\partial^2 W}{\partial S^2}\Big|\sim b^2 |\Lambda|\sim b|W_0|/s$ where use is made of eq.~(\ref{condensescale}). For $s\sim 1$ and $b\sim {\cal O}(10)$,
the $S$-modulus mass can be larger than the additional mass of order $|W_0|$ coming from the hidden brane F-term at the vacuum.
Furthermore, in the presence of the hidden brane F-term, there are changes to the soft mass terms in the minimization conditions for scalars, eqs.~(\ref{minx}) and (\ref{minvarphi}). However, since the mass corrections are still of order $|W_0|$, the minimization of $X$
and $\varphi$ scalars are not altered much.
In Fig.~\ref{plot_s_fzero}, we plot the scalar potential for the real part of the $S$-modulus, before and after the F-term uplifting potential is included. We also show in Fig.~\ref{plot_3d} that the scalar potential for the $S$ modulus and the $X$ scalar has a local minimum as determined approximately by the SUSY conditions.

Finally, after taking into account the hidden brane F-term, we consider the $T$-modulus stabilization and the $U(1)_R$ D-term.
Ignoring ${\hat F}_{X}, {\hat F}_\varphi$, ${\hat F}_S$ and the $U(1)_R$ D-term in the vacuum energy,
the vanishing vacuum energy condition, $V_0\simeq V_F\simeq 0$, gives $\frac{M^2_P e^{|X|^2}}{s}|{\hat F}_{Q'}|^2\simeq 2m^2_{3/2}$.
Then, from eq.~(\ref{dtermdetermine}), the $U(1)_R$ D-term becomes
\be
D_R\simeq -\frac{m^2_{3/2}}{g_R}\Big(1+\frac{1}{2}r_X|X|^2\Big)^{-1}. \label{dtermsimple}
\ee
Therefore, for $r_X|X|^2,r_\varphi|\varphi|^2\ll 1$ 
and when the gravitino mass is much smaller than the $U(1)_R$ gauge boson mass of order $g_R M_P$,
the minimum of the $T$-modulus is shifted from the one determined by the flux in eq.~(\ref{dtermonly}) as
\be
t\simeq 1-\frac{m^2_{3/2}}{2g^2_R M^2_P}-\frac{1}{2}r_X|X|^2-\frac{1}{2}r_\varphi |\varphi|^2.
\ee

\subsection{The scalar soft masses}

The expansion of the K\"ahler potential with respect to the visible sector chiral superfield $Q_i$ gives 
\be
K=K_0(\Phi^a,\Phi^{a\dagger}) +Z_i(\Phi^a,\Phi^{a\dagger}) Q^\dagger_i Q_i
\ee
where $\Phi_a=(S,T,Q',\varphi,X)$ and
\be
Z_i=\Big(\frac{1}{2}(T+T^\dagger)-Q^{'\dagger}Q'-\varphi^\dagger\varphi\Big)^{-1}.\label{zcoeff}
\ee
Here we note that a possible coupling between $Q_i$ and $X$, $\xi_iX^\dagger X Q^\dagger_i Q_i$, in the K\"ahler potential, would induce the soft mass in the presence of nonzero F-terms for $X$.
However, as discussed in the previous section, we can ignore the F-terms for $X$.
Equivalently, we can make an expansion of the superconformal factor $\Omega=-3 e^{-K/3}$ as follows,
\be
\Omega=-3 e^{-K_0/3}+Y_i(\Phi^a,\Phi^{a\dagger}) Q^\dagger_i Q_i
\ee
with
\bea
Y_i=\Big(\frac{1}{2}(S+S^\dagger)\Big)^{\frac{1}{3}}\Big(\frac{1}{2}(T+T^\dagger)-Q^{'\dagger}Q'-\varphi^\dagger \varphi\Big)^{-\frac{2}{3}}.\label{ycoeff}
\eea

In the presence of the SUSY breaking on the hidden brane, we determine the soft scalar mass on the visible brane.
When $Q_i$ and $Q'$ vanish and $F^{Q_i}=0$, using eq.~(\ref{scalarmass}) with (\ref{zcoeff}) or (\ref{ycoeff}), 
we get the general formula for the scalar soft mass as
\bea
m^2_i &=&\frac{1}{M^2_P}V_F+m^2_{3/2}-\frac{|F^T_0|^2}{4t^2}
-\frac{|F^{Q'}|^2}{t}-\frac{|F^{\varphi}|^2}{t}+\Big(-\frac{2}{t}+r_i\Big)g_R D_R \label{softmassgen}
\eea
where $F^T_0$ is given in eq.~(\ref{ft0}).
Now using the $T$-modulus minimization condition (\ref{mint}) and $\frac{|F^T_0|^2}{4t^2}=m^2_{3/2}$, 
we simplify the expression for the scalar soft mass as
\be
m^2_i=\Big(r_i-\frac{1}{t}r_{\varphi}|\varphi|^2\Big)g_RD_R. \label{softmasssimple}
\ee 
Therefore, we find that the $U(1)_R$ D-term is a dominant source for the soft masses in the visible brane.
After using eq.~(\ref{dtermsimple}) in eq.~(\ref{softmasssimple}), for $r_\varphi|\varphi|^2\ll 1$ and $r_X|X|^2 \ll 1$, 
we obtain the scalar soft mass as
\be
m^2_i\simeq -r_i m^2_{3/2}.\label{scalarfinal}
\ee
Consequently, the positive scalar mass squared requires $r_i<0$.
This result agrees with the one obtained in Ref.~\cite{lee} 
where the matter VEVs of the superpotential were assumed not to affect the soft masses.

\section{The $U(1)_R$ phenomenology}

In this section, by using the result of the previous section, we present the detailed soft mass terms 
in the $U(1)_R$ anomaly-free model of section 3.3.
Moreover, we consider the phenomenological implication of the $U(1)_R$ mediation.
We derive the low energy SUSY spectrum and discuss on the constraints
coming from electroweak symmetry breaking, Higgs mass bound from LEP and dark matter relic density from WMAP.

\subsection{The $R$-charges and the scalar soft masses}

From eq.~(\ref{scalarfinal}) with $R$-charges (\ref{scalarrcharge}), we obtain the soft masses for the MSSM scalar fields as
\bea
m^2_{\tilde l}&=&\Big(3\tilde{q}+\frac{16}{3}\Big)m^2_{3/2}, \quad m^2_{\tilde e}=\Big(\frac{3}{7}\tilde{q}+ \frac{26}{21}\Big)m^2_{3/2}, \quad m^2_{\tilde{u}}=-\Big(\frac{17}{7}\tilde{q}+\frac{18}{7}\Big)m^2_{3/2},  \nonumber \\
m^2_{\tilde{d}}&=&\Big(\frac{31}{7}\tilde{q}+\frac{46}{7}\Big)m^2_{3/2}, \quad m^2_{\tilde{h}_d}= -\Big(\frac{24}{7}\tilde{q}+\frac{60}{7}\Big) m^2_{3/2}, 
\quad m^2_{\tilde{h}_u}=\Big(\frac{24}{7}\tilde{q}+\frac{4}{7}\Big)m^2_{3/2}. \label{spectra}
\eea
Then, when the doublet squark $R$-charge lies in the following range, 
\be
-\frac{46}{31} < \tilde{q} < -\frac{18}{17},
\label{range_tildeq}
\ee
we can have all squarks and leptons squared masses to be positive.
This corresponds numerically to $-1.48387<{\tilde q}<-1.05882$.
In this $R$-charge range, the R-charges of the scalar Higgs doublets are
\be
\frac{108}{31}<{\tilde h}_d<\frac{84}{17}, \quad\quad \frac{52}{17}<{\tilde h}_u<\frac{140}{31}.
\ee
Therefore, the soft mass squareds of the scalar Higgs doublets are negative.

\subsection{The gaugino masses}

In the presence of the brane-localized Green-Schwarz term (\ref{GSterm}), the gaugino masses get additional 
corrections due to the $U(1)_R$ anomalies.
Including the anomaly mediation contributions, the general formula for the gaugino masses\footnote{See Ref.~\cite{gauginocode} for the gaugino masses in various schemes of SUSY breaking and mediation.} is
\be
M_a=F^I\partial_I \ln({\rm Re}f_a)+\frac{b_ag^2_a}{8\pi^2}\frac{F^C}{C_0}
\ee
where the compensator superfield is $C=C_0+\theta^2 F_C$ and $b_a=(\frac{33}{5},1,-3)$ are the beta function coefficients in the MSSM.
Here we note that the F-term of the compensator superfield is related to the F-terms of other chiral superfields as
\bea
\frac{F^C}{C_0}=\frac{C^{*2}_0}{C_0}e^{K/2}W^\dagger +\frac{1}{3}K_I F^I.
\eea
Thus, using $k_a=\frac{C_a}{2\pi^2\delta_{GS}}=\frac{C_a}{16\pi^2g_R}$, the gaugino masses become
\bea
M_a=\frac{C_ag^2_a}{16\pi^2 g_R}F^T+\frac{b_ag^2_a}{8\pi^2}\frac{F^C}{C_0}.\label{gauginomass}
\eea
By using $F^T= 2t m_{3/2}\simeq 2m_{3/2}$ and $C_ag^2_a=-9g^2_{\rm GUT}\simeq -\frac{9}{2}$ at the GUT scale, which is obtained from eq.~(\ref{effgaugecoupling}) and eq.~(\ref{universalrel}) with eq.~(\ref{anomfix}),
the $U(1)_R$-anomaly contribution at the GUT scale becomes $M_a\simeq -\frac{9}{16\pi^2 g_R}m_{3/2}$.

Here we note that the effective $U(1)_R$ gauge coupling $g_R$ is given by
\be
\frac{1}{g^2_R} = \frac{V}{g^2} ({\rm Re}S)= \frac{4\lambda \pi}{g^4 M^4_*}({\rm Re}S) 
\ee
where $g$ is the bulk $U(1)_R$ gauge coupling, $V\equiv \lambda \pi r^2_0$ with $r^2_0=4/(g^2M^4_*)$ is the volume of the extra dimensions with $\lambda$ being a deficit angle parameter.
Thus, for both $\lambda$ and ${\rm Re}S$ of order 1, we can get the relation, $g_R\simeq g^2 M^2_*/\sqrt{4\pi}$.
Since the $U(1)_R$ gauge boson mass is given by $M_R=2\sqrt{2}g_R M_P\simeq 4.8\sqrt{2}g_R\times 10^{18}$ GeV, it can be of order the 4D GUT scale for $g_R\simeq 10^{-2}$, which corresponds to $g M_*\simeq 0.2$. 
When $g_R\lesssim \frac{9}{16\pi^2}\simeq 0.057$, which corresponds to $gM_*\lesssim 0.92$, we find $|M_a|\gtrsim m_{3/2}$. 
On the other hand, for $gM_*\sim 1$, $g_R\sim 1/\sqrt{4\pi}$, so
the gaugino mass becomes $M_a\sim -0.2 m_{3/2}$. 
Henceforth, we take $g_R$ to be a free parameter that fixes the gaugino mass.

\subsection{Other soft mass terms}

The supersymmetric action for the brane-localized MSSM chiral superfields coupled to the moduli superfields and a singlet superfield $N$ is
\be
{\cal L}_{\rm vis}=\int d^4\theta C^\dagger C \Big(-3e^{-K_0/3}+Y_i Q_i^\dagger Q_i+Y_N N^\dagger N\Big)
+\Big(\int d^2\theta C^3 W_1(Q_i,N)+{\rm h.c.}\Big)
\ee
where the superpotential is
\be
W_1=\frac{1}{6}\lambda_{ijk}Q_iQ_j Q_k+\lambda_N N^k H {\bar H}
\ee
with $k=-\frac{6}{r_N}$ for $r_N$ being a negative $R$-charge of the singlet superfield $N$.
Here we note that $N^{-k/3}$ term respecting the $U(1)_R$ symmetry can be introduced in the superpotential in order to avoid a $U(1)$ Peccei-Quinn symmetry which would result in a phenomenologically unaccepable axion when the fields get nonzero VEVs.
Then, without anomaly mediation contributions, the full Lagrangian for the soft mass parameters is 
\be
{\cal L}_{\rm soft}=-m^2_i |{\hat Q}_i|^2-\Big(\frac{1}{2}M_a\lambda^a\lambda^a+\frac{1}{6}A_{ijk}y_{ijk}{\hat Q}_i{\hat Q}_j{\hat Q}_k
+\mu B {\hat H}{\hat {\bar H}}+{\rm h.c.}\Big)
\ee
where 
\bea
m^2_i&=&r_i g_R D_R, \\
M_a&=&\frac{C_ag^2_a}{16\pi^2 g_R}F^T, \\
A_{ijk}&=&-F^I\partial_I\ln\bigg(\frac{\lambda_{ijk}}{Y_iY_jY_k}\bigg)=F^I\partial_I\ln(Y_p Y_q Y_r),
\eea
and the supersymmetric Higgs mass parameter $\mu$ and the corresponding soft mass $B$ are given by 
\bea
\mu &=& \frac{\lambda_N\langle N^k\rangle}{C^{k-1}_0 \sqrt{Y^k_N Y_H Y_{\bar H}}},\\
B &=& -F^I\partial_I\ln\bigg(\frac{\lambda_N N^k}{C^{k-1}Y_N Y_H Y_{\bar H}}\bigg)=F^I\partial_I\ln(C^{k-1}Y_N Y_H Y_{\bar H}).
\eea
Here $y_{ijk}$ are the Yukawa couplings for the canonically normalized superfields, ${\hat Q}_i$, related to the original Yukawa couplings as $\lambda_{ijk}=\sqrt{Y_iY_jY_k}y_{ijk}$. 
Therefore, using $Y_i=Y_H=Y_{\bar H}=Y_N$ given in eq.~(\ref{ycoeff}), 
we obtain the soft mass parameters and the $\mu$ term at the GUT scale as
\bea
m^2_i&\simeq&-r_i m^2_{3/2},
\label{soft_terms_msq}  \\
M_a&\simeq& \frac{C_ag^2_a}{8\pi^2 g_R} m_{3/2}\simeq -\frac{9}{16\pi^2 g_R}m_{3/2},
\label{soft_terms_M}  \\
A_{ijk}&=&3\Big(\frac{F^S}{6s}-\frac{F^T}{3t}\Big)\simeq -\frac{F^T}{t}= -2m_{3/2}, \quad {\rm for}\ {\rm any} \ i,j,k,
\label{soft_terms_A} \\
\mu &=&\frac{\lambda_N\langle N^k\rangle}{(s^{\frac{1}{3}}t^{-\frac{2}{3}})^{\frac{1}{2}k+1}},
\label{soft_terms_mu} \\
B&=&(k-1)\frac{F^C}{C_0}+3\Big(\frac{F^S}{6s}-\frac{F^T}{3t}\Big)\simeq \frac{2}{3}(k-4)m_{3/2}
\label{soft_terms_B}
\eea
where we used $\frac{F^C}{C_0}\simeq m_{3/2}-\frac{F^T}{6t}= \frac{2}{3}m_{3/2}$
and $C_ag^2_a=-9g^2_{\rm GUT}\simeq -\frac{9}{2}$ at the GUT scale. 
Here we note that the anomaly mediation contribution to the gaugino mass is also ignorable 
for a small $g_R$ and we assumed that the SUSY breaking from the singlet $N$ is negligible.

\subsection{Low energy spectrum and phenomenology}

\begin{figure}[!t]
  \begin{center}
  \begin{tabular}{c c}
   \includegraphics[width=0.5\textwidth]{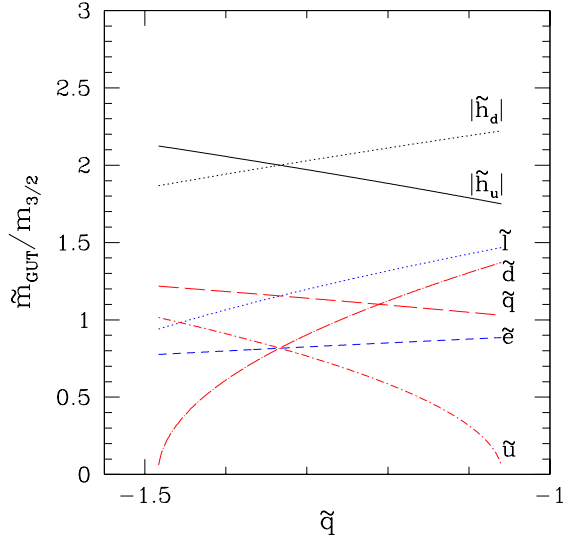}
&
   \end{tabular}
  \end{center}
 \caption{The soft masses ${\tilde m}_{\rm GUT}$ for sparticles at the GUT scale with a varying $\tilde{q}$.
For the Higgs masses, we plot ${\tilde m}_{\rm GUT}=\sqrt{|m^2_{\tilde{h}_{u,d}}|}$.}
\label{GUT_tildeq}
\end{figure}

We have shown in the previous section that all the scalar soft masses are determined by the $R$-charge of the doublet squark($\tilde q$) and the gravitino mass($m_{3/2}$) while the gaugino mass($M_{1/2}$) is determined by the $U(1)_R$ gauge coupling. Thus, in our $U(1)_R$ mediation model, there are five free parameters determining the soft mass parameters at low energy:
\be
m_{3/2}, \quad M_{1/2}, \quad \tilde q, \quad \tan\beta, \quad  {\rm sign}(\mu). 
\ee
Here we assume that the $\mu$ term and the $B$ term are being used in minimizing the scalar potential of the visible sector for electroweak symmetry breaking.

In \fig{GUT_tildeq}, we show the soft scalar masses at the GUT scale 
for a varying $\tildeq$ according to \eq{spectra}.
With a decreasing $|\tildeq|$, the masses of sleptons and down-type squarks
increase while those of squark doublets and up-type squarks decrease.
Especially $m^2_{\tilde{u}}$ and $m^2_{\tilde{d}}$ start being negative at either boundary value of 
$\tildeq$, which gives the limit on the range of $\tildeq$ as was seen in
 \eq{range_tildeq}.

For the Higgs doublets, their mass squareds are always negative 
in the given $\tildeq$ range in \fig{GUT_tildeq}, therefore we plot $\sqrt{|m^2_{\tilde{h}_{u,d}}|}$.
The absolute masses of $\tilde{h}_d$ ($\tilde{h}_u$)
are increasing (decreasing) as $|\tildeq|$ decreases.
In particular, for $\tildeq=-4/3$, 
we get the $R$-charges as ${\tilde h}_u={\tilde h}_d=4$, 
${\tilde l}=-\frac{4}{3}$ and ${\tilde e}={\tilde u}={\tilde d}=-\frac{2}{3}$.
In this case, we get the same scalar soft masses for Higgs doublets 
and almost degenerate soft masses for squarks and sleptons as 
$m^2_{{\tilde h}_d}=m^2_{{\tilde h}_u}=-4m^2_{3/2}$, 
$m^2_{\tilde q}=m^2_{\tilde l}=\frac{4}{3}m^2_{3/2}$ and 
$m^2_{\tilde e}=m^2_{\tilde u}=m^2_{\tilde d}=\frac{2}{3}m^2_{3/2}$.
This case is very similar to the NUHM1 model 
with negative Higgs mass-squareds in~\cite{Baer:2005bu}.

\begin{figure}[!t]
  \begin{center}
  \begin{tabular}{c c}
   \includegraphics[width=0.5\textwidth]{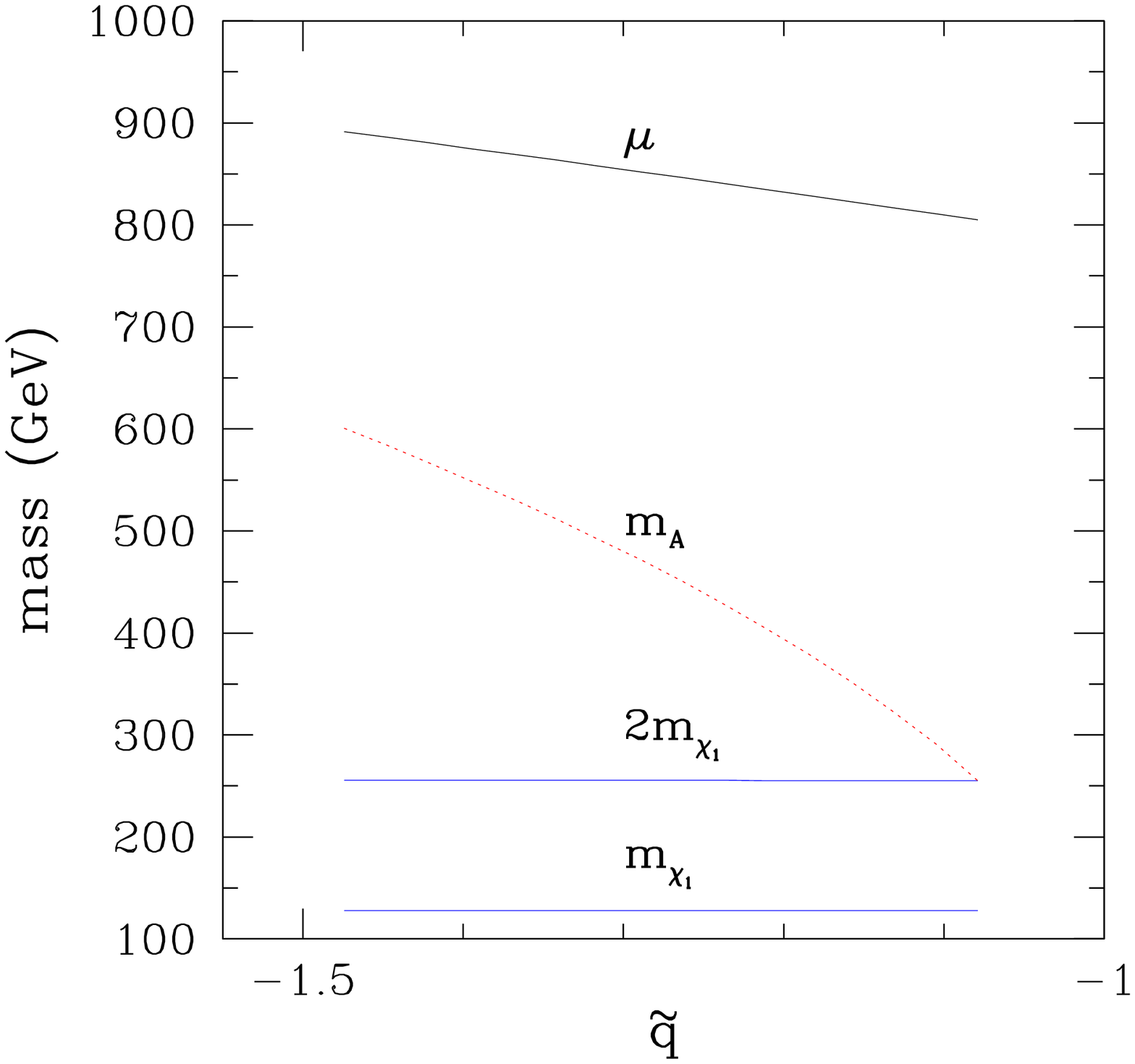}
&
   \includegraphics[width=0.5\textwidth]{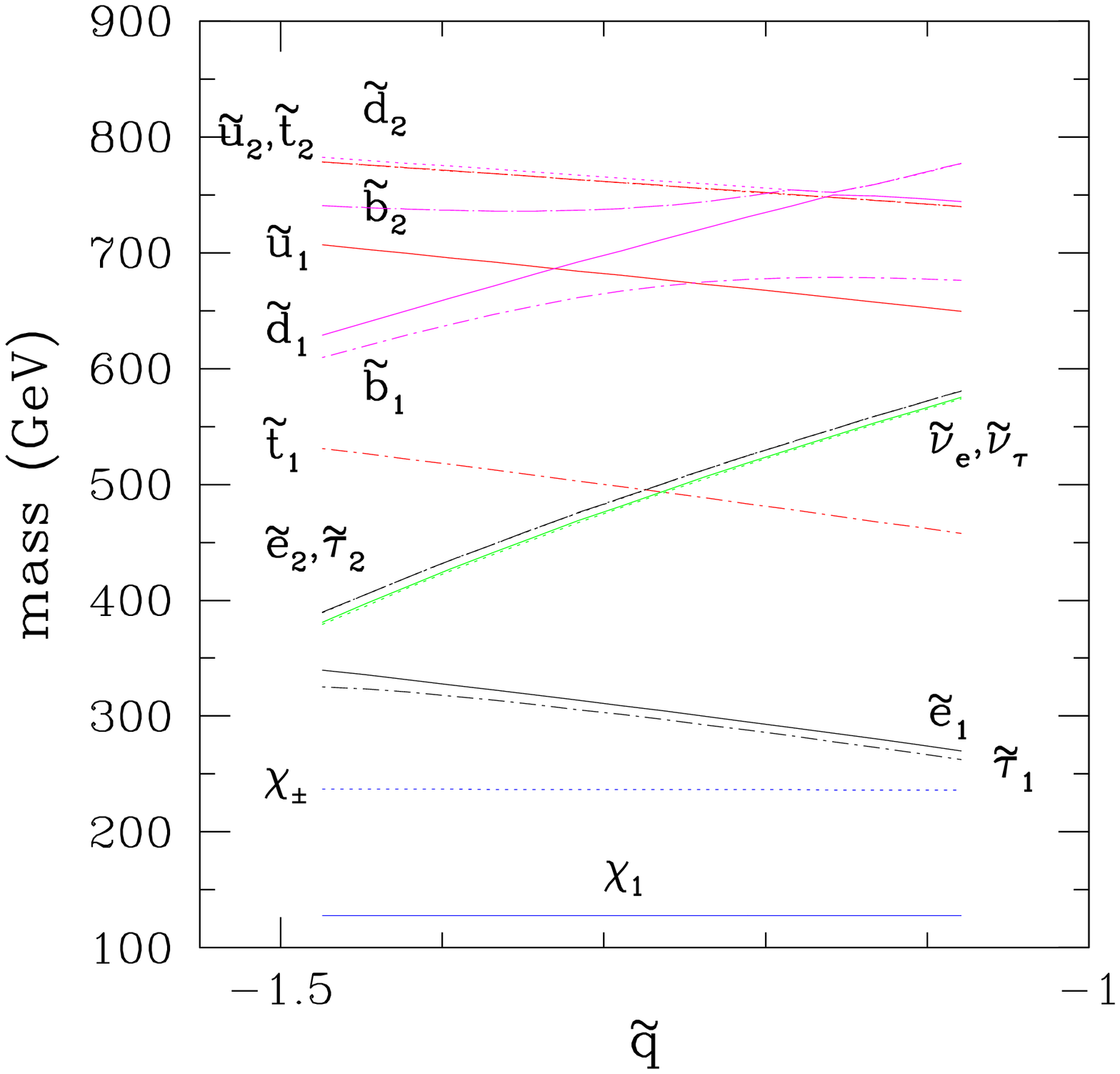}
   \end{tabular}
  \end{center}
 \caption{The particle masses versus $\tilde{q}$ at low energy in the model
for $m_{3/2}=360 \gev$, $M=310\gev$, $\tan\beta=10$ with $\mu>0$ and 
$m_t=172.7 \gev$. In the left window, neutralino (blue solid), 
chargino (blue dotted) and $\mu$ (black solid) are shown.
In the right window, up squarks (red), 
 down squarks (magenta), sleptons (black), sneutrinos (green) are shown.}
\label{mu_tildeq}
\end{figure}

In Fig.~\ref{mu_tildeq} we show the variation in masses at low
energy scale. In Fig.~\ref{mu_tildeq} (left)
we plot the values of $\mu$, $m_A$, $m_{\chi}$ and $2m_\chi$ versus
$\tildeq$ while fixing $\mgra=360\gev$, $\mgauge=310\gev$ with $\tb=10$ and
$\mu>0$.
The parameter $\mu$ is large around $850\gev$ and the behavior of
$\mu$ vs. $\tildeq$ can be understood from the relation 
$\mu^2\sim -m_{{\tilde h}_u}^2$ at low energy 
for moderate to large values of $\tb$ and 
$|m_{{\tilde h}_u}|\gg M_Z$~\cite{Baer:2005bu}.
The value of $\mu$ becomes slightly small 
when $\tildeq$ becomes large, since $|m_{{\tilde h}_u}^2|$ decreases.
The tree level pseudoscalar Higgs mass $m_A$ is given by~\cite{Baer:2005bu}
\dis{
m_A^2=\mHu^2+\mHd^2+2\mu^2\simeq \mHd^2-\mHu^2.
}
Since both $\mHu^2$ and $\mHd^2$ are negative, they can cancel against $2\mu^2$
term resulting in pseudoscalar Higgs mass small.
For a smaller $|\tilde q|$, e.g. ${\tilde q}=-1.1$, $\mHu^2 > \mHd^2$ 
at GUT scale, 
thus $m^2_A$ becomes easily negative at low energy after RGE running.  
In particular, when $m_A\sim2m_{\chi_1}$ with small value $|\tilde q|$ 
before it could reach the negative values, 
 neutralinos in the early universe may annihilate efficiently 
through heavy Higgs resonances, called `the A-annihilation funnel',
so that the right relic density of neutralino dark matter can be obtained.
Since the Higgs mass-squared is negative the A-annihilation funnel occurs
even for low $\tan\beta$ values~\cite{Baer:2005bu}.
The neutralino mass is relatively invariant for the change in $\tildeq$,
since it is closely related to the gaugino mass as $m_{\chi_1}\sim 0.4 M_{1/2}$.
In the right of Fig.~\ref{mu_tildeq}, we show the different sparticle masses with 
$\tildeq$ for the same mass parameter choices as in the left.

\begin{figure}[!t]
  \begin{center}
  \begin{tabular}{c c}
   \includegraphics[width=0.5\textwidth]{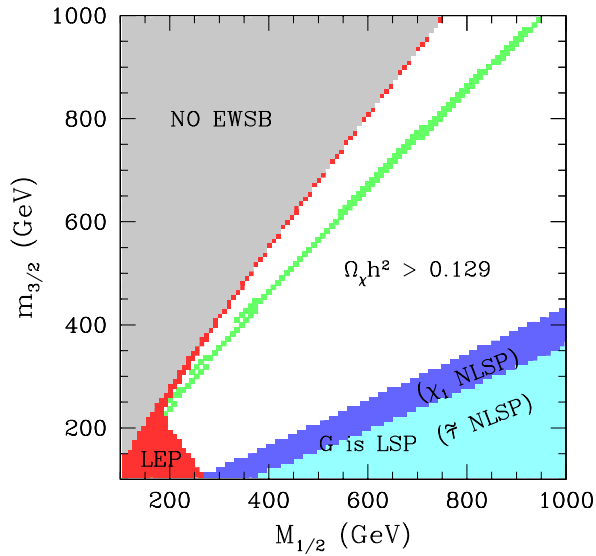}
&
   \end{tabular}
  \end{center}
 \caption{The scan on the plane of $(M_{1/2},m_{3/2})$ with $\tilde{q}=-1.1$,
$\tan\beta=10$ and $\mu>0$.
The black region is excluded due to unsuccessful EWSB (upper left corner). 
The red region is disfavored
by the LEP constraints on chargino and Higgs mass $m_{\chi\pm} > 104 \gev$ and
$m_{h^0} > 114.4 \gev$. In the dark-blue region, gravitino is LSP and neutralino is NLSP.
In the light-blue region, gravitino is LSP and stau is NLSP. 
In the green region, the relic density of neutralino is less than the WMAP upper bound.}
\label{scan_tb10}
\end{figure}

In \fig{scan_tb10} we show the $\mgauge$ vs $\mgra$ parameter space
with $\tb=10$, $\mu>0$, $m_t=172.7\gev$ and $\tildeq=-1.1$.
For obtaining this we used the Fortran package SUSPECT~\cite{Djouadi:2002ze}
for low energy spectrum and DarkSusy~\cite{Gondolo:2004sc} for dark matter relic density.
The black regions are excluded by lack of REWSB (left upper corner).
The red regions are excluded by the LEP2 constraint that $m_{\chi_\pm}>104\gev$
and $m_{h^0}>114.4\gev$.
In the blue region, the gravitino is LSP and neutralino is NLSP. 
On the other hand, in the light-blue region, the gravitino is LSP while stau is NLSP.
In the remaining parameter space, we implemented the relic density of 
neutralino in the green region denoting the upper bound from
WMAP, $\oh<0.129$~\cite{Komatsu:2008hk}.
We can see the narrow green band in the upper middle of the parameter plane,
which is the A-annihilation funnel. 
This was also shown in~\cite{Baer:2005bu}.

\begin{figure}[!t]
  \begin{center}
  \begin{tabular}{c c}
   \includegraphics[width=0.5\textwidth]{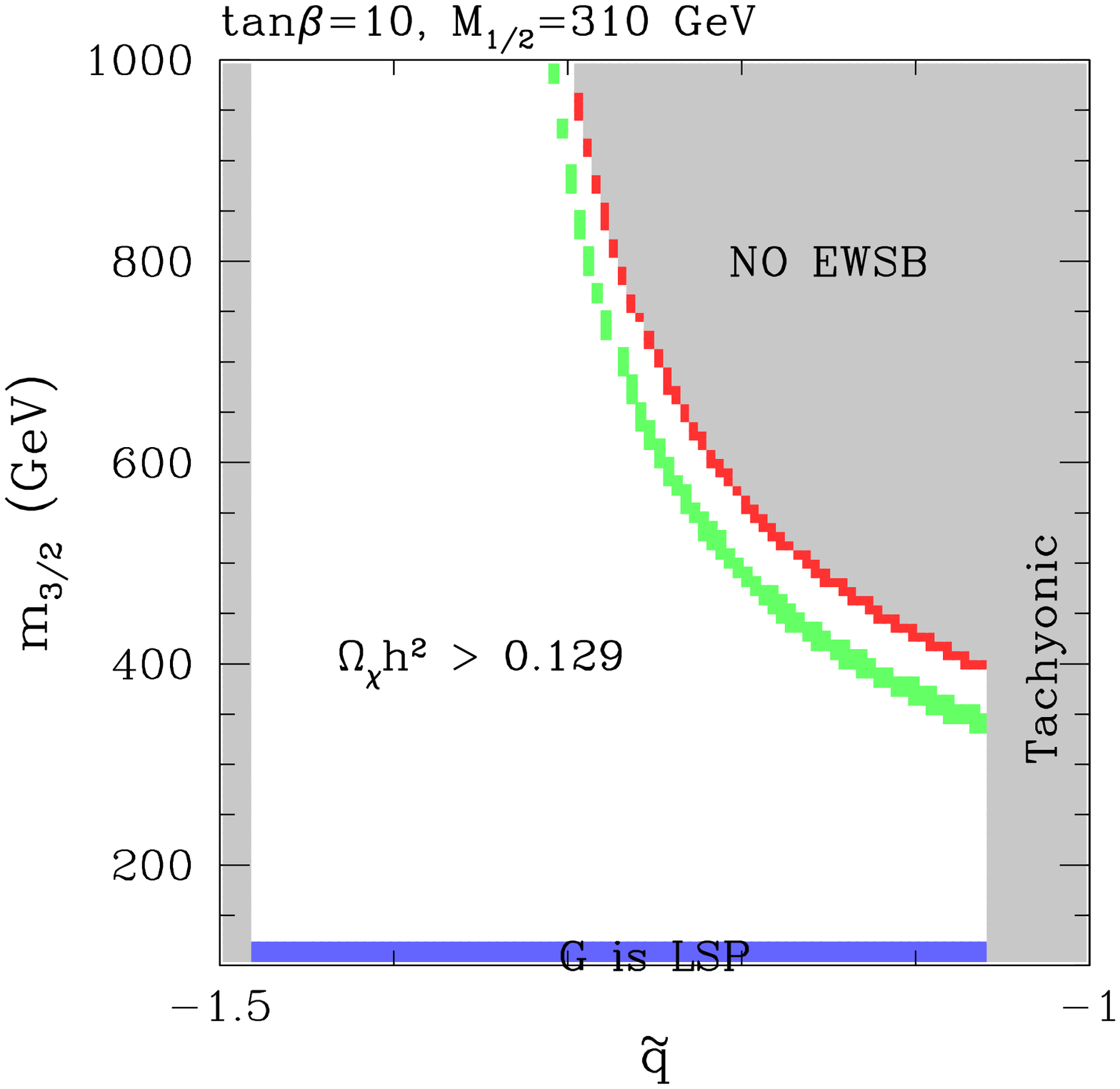}
&
   \includegraphics[width=0.5\textwidth]{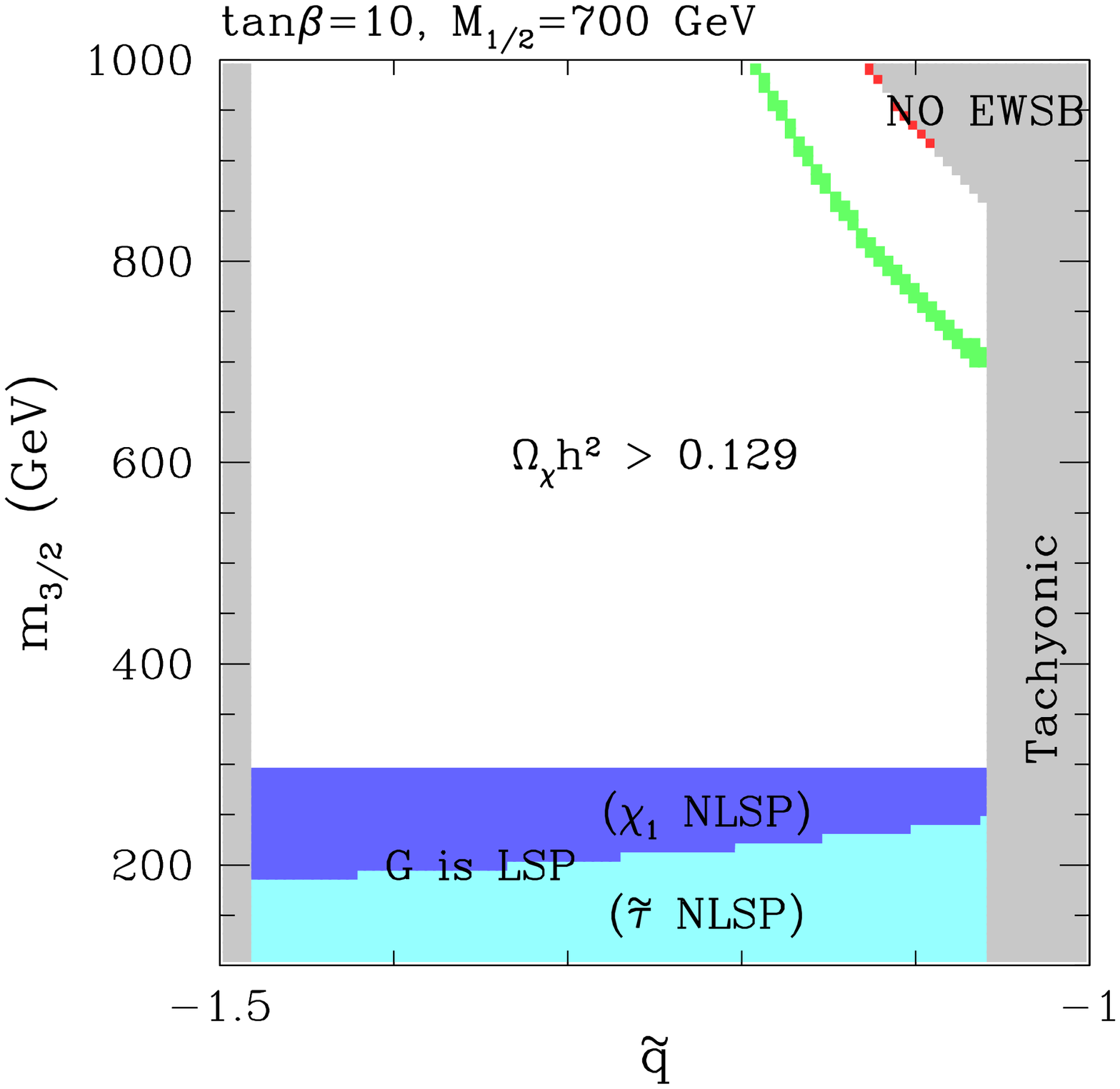}
   \end{tabular}
  \end{center}
 \caption{The gravitino mass vs $\tilde q$ 
with $\tan\beta=10$ and $\mu>0$
for $M_{1/2}=310 \gev$ (Left) and  $M_{1/2}=700 \gev$ (Right). 
The black region are excluded due to tachyonic charged sfermions 
and unsuccessful EWSB. The red region is disfavored by LEP constraints.
In the dark-blue region, gravitino is LSP and neutralino is NLSP. 
In the light-blue region, gravitino is LSP and stau is NLSP. 
In the green region, the neutralino relic density is less than 
the WMAP upper bound.}
\label{scanMm0}
\end{figure}

In \fig{scanMm0}, we also plot the $\tilde q$ vs 
gravitino mass parameter space 
with $\tan\beta=10$, $\mu>0$, $m_t=172.7\gev$ and $M_{1/2}=310$ GeV (Left)
and  $M_{1/2}=700$ GeV (Right).
For an increasing $|\tildeq|$ the available parameter space becomes larger
since  $\mHd^2 > \mHu^2$ as said before.
However the A-funnel region tends to disappear since the pseudo scalar Higgs mass
becomes much larger than the twice of neutralino mass.
Then there is no viable region for neutralino LSP with right relic density.
With a increasing $\tb$, the parameter space shrinks and the A-funnel region also becomes smaller.
This is because the larger bottom type quark Yukawa coupling leads to a larger
 negative running to $m^2_{{\tilde h}_d}$ so that $m^2_A$ gets negative.

The characteristic property of our $U(1)_R$-mediation model is that 
in the stau-neutralino coannihilation region, which is within the dark-blue region of \fig{scan_tb10}, 
gravitino becomes lighter than neutralino.
So, the neutralino DM is possible only at the A-annihilation funnels
mentioned above.
However the blue region with either neutralino (dark blue) 
or stau (thin blue) NLSP is compatible since gravitino is LSP
and it can be a dark matter component with 
thermal and nonthermal production.
However the decay of NLSP can produce electromagnetic
and/or hadronic particles into the expanding plasma
when they decay in the early universe, and this
can change the light element abundances resulting in the severe problems
with observation~\cite{Cyburt:2002uv,Kawasaki:2004yh,Kawasaki:2004qu}.
Neutralino NLSP is very sensitive to Big Bang Nucleosynthesis (BBN) 
constraint and very difficult to be viable NLSP~\cite{Feng:2003uy}.
However stau NLSP is less sensitive to the BBN constraint and
gives a good reason for the gravitino to be LSP dark matter~\cite{Ellis:2003dn,Roszkowski:2004jd,Jedamzik:2005dh,Cerdeno:2005eu,Pradler:2006hh,kai0}.
Considering the bound state effects of charged particle during BBN epoch,
the lifetime of stau NLSP is constrained to be smaller than about 
$5\times10^3$sec~\cite{Pospelov:2006sc,Pradler:2007is,Bailly:2008yy}.
The problem of the light neutralino/stau NLSP may be evaded by 
making them heavier than 1-10 TeV  and/or by 
introducing a small breaking of $R$-parity 
as discussed in Ref.~\cite{Buchmuller:2007ui}. 
In the case with substantial left-right stau mixing,
the BBN problem with stau NLSP can be also evaded 
because the annihilation into Higgs bosons reduces the thermal relic density of staus\cite{kai}.

\begin{table}[h]
\centering
\begin{tabular}{|c|c|c|c|c|}
\hline
 & P1 &P2 &P3 &P4\\
\hline
 $\mgra$    &360 &756& 175& 250\\
 $M_{1/2}$  &310 &700& 500& 800\\
 $\tildeq$  & -1.1&-1.1&-1.1 & -1.1\\
 $\tb$& 10&10&10&10\\
 \hline 
 $\mu$& 810& 170& 744 & 111\\
 $m_{h^0}$& 115& 120& 116 & 119\\
 $m_A$& 284& 626& 715 &1100 \\
 $m_{H^0}$& 284 & 626&  715& 1100\\
 $m_{H^\pm}$& 295 & 631 & 720& 1103\\
 \hline
 $m_{\chi_1}$  &  127&  299&  207&  339\\  
 $m_{\chi_2}$  &  246&  572&  393&  641\\    
 $m_{\chi_3}$  & 806& 1690& 747& 1117\\  
 $m_{\chi_4}$  &  811&  1691& 757& 1124\\  
 $m_{\chi^\pm_1}$&  246&  572&  393&  641\\  
 $m_{\chi^\pm_2}$& 811& 1691& 757&  1124\\  
 $m_{\tilde{g}}$& 754& 1600& 1148& 1772\\ 
 \hline
 $m_{\tilde{u}_1}$& 677& 1420& 1010& 1549\\  
 $m_{\tilde{t}_1}$& 492& 1095&  781& 1228\\  
 $m_{\tilde{d}_1}$& 768& 1612& 1027& 1568\\  
 $m_{\tilde{b}_1}$& 710& 1495&  975& 1498\\  
 $m_{\tilde{e}_1}$& 274&  581&  224&  342\\  
 $m_{\tilde{\tau}_1}$&267&570&  215&  333\\ 
\hline
 $\oh$& 0.1115& 0.1099& $\chi_1$ NLSP& $\tilde{\tau}$ NLSP\\
\hline
 LSP&  $\chi_1$& $\chi_1$&Gravitino&Gravitino\\
\hline  
\end{tabular}
 \caption{All masses are in GeV. P1, P2: A-annihilation funnel. 
P3: Gravitino LSP with neutralino NLSP,
P4: Gravitino LSP with stau NLSP.}
 \label{table:spectra}
\end{table}

In Table~\ref{table:spectra}, we show four examples of low-energy
spectrum with neutralino LSP DM (P1, P2) and gravitino LSP (P3, P4).
In the point (P1, P2) the WMAP constraint of relic density is achieved
through the A-annihilation funnel, where $2m_{\chi_1}\sim m_A$.
In the point (P3) the gravitino is LSP and the neutralino is NLSP.
In the point (P4) the gravitino is LSP and the stau is NLSP.
In these latter points the thermally produced gravitino can provide the relic density for dark matter
with reheating temperature $T_R=10^{7-9}$ GeV\cite{Cerdeno:2005eu}.

Finally, some brief remarks on the experimental signature of the $U(1)_R$ mediation are in order.
As discussed before, due to the negative Higgs mass squareds at the GUT scale, it is possible to decrease
the CP-odd($A$) Higgs boson mass even at low $\tan\beta$ values in the A-funnel region for relic density being in accordance with the WMAP result. In P1 with $m_A=284$ GeV, the heavier neutral $H^0$ and $A$ Higgs bosons are light so they should be detected at the LHC via direct $H^0$ and $A$ production followed by $H^0,A\rightarrow \tau{\bar\tau}$ decay\cite{pseudoscalardetec,djouadi}.
Moreover, since the charged $H^\pm$ Higgs boson is also rather light, the gluon fusion process $gb\rightarrow tH^+$ followed by $H^+\rightarrow \tau^+\nu_\tau$ could be observed at the LHC\cite{djouadi}.
At the LHC, squarks and gluinos will be copiously produced due to the large QCD coupling.
Produced squarks decay into quark and the second lightest neutralino $\chi_2$ or the light chargino $\chi^+_1$.
In P1, since $\chi_2$ and $\chi^+_1$ are lighter than sleptons and squarks, $\chi_2$ decays to $Z(h^0)$ and $\chi_1$
while $\chi^+_1$ decays to $W^+$ and $\chi_1$. For three body decays of $\chi_2$ and $\chi^+_1$ into leptons, it would be easier to identify the signal, apart from the limiting factor of the identification of $\tau$'s in the final state.
On the other hand, produced gluinos decay dominantly to stop and top quark. So, after cascade decays of gluinos, 
the gluino pair production leads to 4 jets plus transverse missing energy.

\section{Conclusion}

We have considered a 4D effective supergravity with gauged $U(1)_R$ which was recently derived from a supersymmetric flux compactification with codimension-two branes in 6D chiral gauged supergravity. 
We presented a $U(1)_R$ anomaly-free model where the MSSM fields are the only SM non-singlets
and obtained the consistent $R$-charges of the MSSM fields.
The SM-$U(1)_R$ anomalies are cancelled by a $T$-modulus dependent Green-Schwarz counterterms localized on the brane.

Stabilization of a remaining modulus, the $S$-modulus, needs an introduction of a bulk gaugino condensate that depends on the $S$-modulus because of the nontrivial bulk gauge kinetic term. The effective superpotential for the gaugino condensate is manifestly $U(1)_R$-invariant and a constant term in the superpotential is introduced from another sector that breaks the $U(1)_R$ symmetry spontaneously satisfying a local SUSY condition. For a nonzero superpotential VEV, the potential minimization with respect to the $T$-modulus leads to a $U(1)_R$ D-term of order the gravitino mass. Consequently, after the stabilization of all the moduli at a vanishing vacuum energy, we find that the $U(1)_R$ D-term can be a dominant source of soft masses for scalar fields with nonzero $R$-charge. 

We found that there is a parameter space of the $R$-charges which is compatible with the $U(1)_R$ anomaly cancellation and at the same time leads to positive soft mass squareds for all squarks and sleptons. 
At the GUT scale, the scalar soft masses are family-independent but they are not universal. 
Moreover, in the presence of a nonzero $T$-modulus F-term, the brane-localized anomaly corrections give rise to nonzero universal gaugino masses at the GUT scale.
For a reasonably small $U(1)_R$ gauge coupling, the gaugino masses can be of order the gravitino mass.
Consequently, for the phenomenology of $U(1)_R$-mediated SUSY breaking, we obtained the low energy superparticle spectrum and constrained the model parameters by considering correct electroweak symmetry breaking and the Higgs mass bound from LEP as well as dark matter relic density from WMAP. 
We have shown that neutralino can be an LSP and it can satisfy the dark matter density bound through the pseudoscalar Higgs annihilation channels. 
We also found that in the stau-neutralino coannihilation region, gravitino is a LSP and it can be a good dark matter 
candidate. In this case, however, because of the decay of NLSP, which is neutralino or stau, 
there is a strong constraint coming from the BBN.
We discussed briefly on the experimental signature of the $U(1)_R$ mediation, in relation to a light pseudoscalar Higgs mass required for explaining the dark matter relic density in the $A$-funnel region.

\section*{Acknowledgments}
We would like to thank Cliff Burgess for discussion.
H.M.L. is supported by the research fund from the Natural Sciences and Engineering Research Council (NSERC) of Canada. 
K.-Y.C. is supported by
the Ministerio de Educacion y Ciencia of Spain under Proyecto Nacional FPA2006-05423 and
by the Comunidad de Madrid under Proyecto HEPHACOS, Ayudas de I+D S-0505/ESP-0346.
K.-Y.C. would like to thank the European Network of Theoretical 
 Astroparticle Physics ILIAS/ENTApP under contract number 
 RII3-CT-2004-506222 for financial support.

\def\theequation{A.\arabic{equation}}
\setcounter{equation}{0}
\vskip0.8cm
\noindent
{\Large \bf Appendix A: The K\"ahler metric and the F-terms}
\vskip0.4cm
\noindent

Setting $V_R$ to zero, the K\"ahler potential only with brane matters takes the following form,
\bea
K&=&-\ln\Big(\frac{1}{2}(S+S^\dagger)\Big) \nonumber \\
&&-\ln\Big(\frac{1}{2}(T+T^\dagger)-\Omega(Q^\dagger,Q)-\Omega'(Q^{'\dagger},Q')\Big).
\label{kahler1}
\eea
Then, the K\"aher metric $K_{I{\bar J}}=\partial_I\partial_{\bar J}K$ ($\Phi^I=Q,Q',T,S$) is given by
\be
K_{I{\bar J}}=\left(
\begin{array}{llll}
\frac{\Omega_{Q{\bar Q}}}{t}+\frac{|\Omega_{Q}|^2}{t^2} \,\,& \frac{\Omega_{Q}\Omega'_{\bar {Q'}}}{t^2} \,\,& -\frac{\Omega_Q}{2t^2} & 0\\
\frac{\Omega'_{Q'}\Omega_{\bar Q}}{t^2} \,\,& \frac{\Omega'_{Q' {\bar {Q'}}}}{t}+\frac{|\Omega'_{Q'}|^2}{t^2} \,\,& -\frac{\Omega'_{Q'}}{2t^2} & 0 \\
-\frac{\Omega_{\bar Q}}{2t^2} \,\,& -\frac{\Omega'_{\bar {Q'}}}{2t^2}\,\, & \frac{1}{4t^2} & 0 \\
0 \,\,& 0 \,\,& 0 & \frac{1}{4s^2}
\end{array}\right)
\ee
where
\be
t=\frac{1}{2}(T+T^\dagger)-2\Omega-2\Omega', \quad s=\frac{1}{2}(S+S^\dagger).
\ee
The inverse metric is
\be
K^{{\bar I}J}=\left(
\begin{array}{llll}
\frac{t}{\Omega_{Q{\bar Q}}}\,\, & 0 \,\,& \frac{2\Omega_Q}{\Omega_{Q{\bar Q}}}t & 0  \\
0 \,\,& \frac{t}{\Omega'_{Q'{\bar {Q'}}}} \,\,& \frac{2\Omega'_{Q'}}{\Omega'_{Q'{\bar {Q'}}}}t & 0 \\
\frac{2\Omega_{\bar Q}}{\Omega_{Q{\bar Q}}}t \,\,& \frac{2\Omega'_{\bar {Q'}}}{\Omega'_{Q'{\bar {Q'}}}}t \,\,&
4t\Big(t+\frac{|\Omega_Q|^2}{\Omega_{Q{\bar Q}}}+\frac{|\Omega'_{Q'}|^2}{\Omega'_{Q'{\bar {Q'}}}}\Big) & 0 \\
0 \,\,& 0 \,\,& 0 & 4s^2
\end{array}\right).
\ee

The F-terms are
\bea
F^T/M_P&=&-4\sqrt{\frac{t^3}{s}}(D_T W)^\dagger+2(\Omega_QF^{Q}+\Omega'_{Q'}F^{Q'}), \label{ftermfort}\\
F^S/M_P&=&-4\sqrt{\frac{s^3}{t}}(D_S W)^\dagger, \\
F^{Q'}/M_P&=&-\sqrt{\frac{t}{s}}\frac{1}{\Omega'_{Q'{\bar {Q'}}}}((D_{Q'}W)^\dagger+2\Omega'_{\bar{Q'}}(D_T W)^\dagger),\\
F^{Q}/M_P&=&-\sqrt{\frac{t}{s}}\frac{1}{\Omega_{Q{\bar Q}}}((D_{Q}W)^\dagger+2\Omega_{\bar{Q}} (D_T W)^\dagger).
\eea
In the text, we use the K\"ahler metric and the F-terms for $\Omega=Q^\dagger Q$ and $\Omega'=Q^{'\dagger}Q'$. 
It is straightforward to generalize to the case with multiple brane chiral multiplets by taking the sum in the F-term for the $T$-modulus, (\ref{ftermfort}).

\def\theequation{B.\arabic{equation}}
\setcounter{equation}{0}
\vskip0.8cm
\noindent
{\Large \bf Appendix B: The scalar potential and the soft masses in 4D supergravity}
\vskip0.4cm
\noindent

We first summarize the scalar potential and the soft masses in the general 4D supergravity
and then apply the result to our example of a gauged $U(1)_R$ supergravity considered in the text.

\section*{B1. The general 4D supergravity}

The 4D supergravity is described by two functions of superfields, the K\"ahler potential $K$, which is a real function, and the superpotential $W$, which is a complex function.
For a given set of the K\"ahler potential and the superpotential, the general scalar potential in 4D supergravity is given by
\be
V_0= V_F + V_D
\ee
where $V_F$, $V_D$ are F-term and D-term potentials, respectively, as follows,
\bea
V_F&\equiv& M^2_P K_{I{\bar J}}F^I F^{J\dagger}-3M^4_P e^{K}|W|^2,\\
V_D&\equiv&\frac{1}{2}({\rm Re}f_a) D^a D^a.
\eea
Here $F^I=-M_P e^{K/2}K^{I{\bar J}} (D_J W)^\dagger$ with $D_I W=\frac{\partial W}{\partial \Phi_I}+\frac{\partial K}{\partial\Phi_I} W$ for $\Phi_I$ being visible sector fields($Q_i$) as well as hidden sector and moduli fields($\Phi_a$). 
$f_a$ is called the gauge kinetic function the real part of which corresponds to the coefficient of the gauge kinetic term.
The D-term is given in a general expression as
\be
D^a=\frac{M^2_P}{{\rm Re}f_a}(-i\eta^I_a\partial_I K+3ir_a)
\ee
with gauge transformations $\delta_a\Phi^I=\eta^I_a(\Phi)$ and $\delta_a W=-3r_a W$.
Here we note that using $\delta_a W=\eta^I_a\partial_I W$, for a nonzero superpotential or gravitino mass,
there is an identity relation between the D-term and the F-term as follows,
\be
D^a=\frac{i}{{\rm Re}(f_a)}\frac{\eta^I_a F_I}{m_{3/2}}M^2_P \label{dtermconsist}
\ee
with $F_I=K_{I{\bar J}}F^{J\dagger}$ and $m_{3/2}=e^{K/2} W$.

Generalizing the general formula\cite{softmass} for the tree-level soft scalar mass obtained in the case without D-terms to 
the case with D-terms included, we obtain the soft mass of a scalar as follows,
\bea
m^2_i &=& \frac{1}{M^2_P}V_F+m^2_{3/2}-F^I F^{J\dagger}\partial_I\partial_{\bar J} \ln Z_i \nonumber \\
&&+K^{Q^\dagger_iQ_i}(({\rm Re} f_a) D^a\partial_i\partial_{\bar i}D^a
+\frac{1}{2}D^aD^a\partial_i\partial_{\bar i}({\rm Re}f_a))  \nonumber \\
&=&\frac{2}{3M^2_P}V_F-F^I F^{J\dagger}\partial_I\partial_{\bar J} \ln Y_i \nonumber \\
&&+K^{Q^\dagger_iQ_i}(({\rm Re} f_a) D^a\partial_i\partial_{\bar i}D^a
+\frac{1}{2}D^aD^a\partial_i\partial_{\bar i}({\rm Re}f_a))
 \label{mostgeneral}
\eea
where use is made of $V_F=V_0-V_D$ and $3m^2_{3/2}M^2_P=-V_0+V_D+M^2_PK_{I{\bar J}}F^IF^{J\dagger}$ in the second line.
Here $Y_i$ and $Z_i=e^{K_0/3}Y_i$ are independent of the visible sector fields $Q_i$, 
and they can be read from the expansion of the K\"ahler potential as follows,
\be
K= K_0(\Phi^a,\Phi^{a\dagger}) + Z_i(\Phi^a,\Phi^{a\dagger}) Q^\dagger_i Q_i,
\ee
where $\Phi^a$ are hidden sector and moduli fields,
or from the expansion of the superconformal factor $\Omega=-3e^{-K/3}$,
\be
\Omega=-3e^{-K_0/3}+Y_i(\Phi^a,\Phi^{a\dagger}) Q^\dagger_i Q_i. 
\ee

\section*{B2. The gauged $U(1)_R$ supergravity}

In the 4D effective supergravity with gauged $U(1)_R$ that we are considering in this paper, 
for the K\"ahler potential (\ref{kahlergen}), using the result in the previous sections, 
we obtain the F-term potential as
\bea
V_F&=&\frac{M^2_P|F^T|^2}{4t^2}+\frac{M^2_P|F^S|^2}{4s^2}+M^2_P|F^{X_i}|^2-\frac{3M^4_P|W|^2}{st} e^{|X_i|^2} \nonumber \\
&&+\sum_{I=Q_i,Q',\varphi}\bigg[\Big(1+\frac{|\Phi_I|^2}{t}\Big)\frac{M^2_P|F^{\Phi_I}|^2}{t}
-\frac{1}{2t^2}(\Phi_I F^T F^{\Phi_I^\dagger} +{\rm h.c.})\bigg] \nonumber \\
&&+\sum_{I,J=Q_i,Q',\varphi,I\neq J}\frac{1}{2t^2}(\Phi_I \Phi^\dagger_J F^{\Phi_J} F^{\Phi_I \dagger} +{\rm h.c.})
\label{ftermpot}
\eea
with $t=\frac{1}{2}(t+t^\dagger)-Q^\dagger_i Q_i -Q^{'\dagger}Q'-\varphi^\dagger\varphi$,
and the D-term potential for the $U(1)_R$ is
\be
V_D=\frac{1}{2}sD^2_R  \label{dtermpot}
\ee
with
\be
D_R=\frac{2g_R M^2_P}{s}\bigg[1-\frac{1}{t}+\frac{1}{2t}(r_i|Q_i|^2+r'|Q'|^2+r_{\varphi}|\varphi|^2)+\frac{1}{2}r_{X}|X|^2\bigg]. \label{scalarsdterm}
\ee
Here the F-terms are given by
\bea
F^T&=&-4M_P e^{\frac{1}{2}|X|^2}\sqrt{\frac{t^3}{s}}(D_T W)^\dagger+2Q^{'\dagger}F^{Q'}
+2\varphi^\dagger F^{\varphi}+2Q^\dagger_i F^{Q_i}, \\
F^{S}&=&-4M_P e^{\frac{1}{2}|X|^2}\sqrt{\frac{s^3}{t}}(D_S W)^\dagger, \\
F^{Q'}&=&-M_P e^{\frac{1}{2}|X|^2}\sqrt{\frac{t}{s}}\,(\partial_{Q'}W+2Q^{'\dagger}\partial_T W)^\dagger, \\
F^{Q_i}&=&-M_P e^{\frac{1}{2}|X|^2}\sqrt{\frac{t}{s}}\,(\partial_{Q_i}W+2Q^{\dagger}_i\partial_T W)^\dagger, \\
F^{\varphi}&=&-M_P e^{\frac{1}{2}|X|^2}\sqrt{\frac{t}{s}}\,(\partial_{\varphi}W+2\varphi^{\dagger}\partial_T W)^\dagger, \\
F^{X}&=&-M_P \frac{e^{\frac{1}{2}|X|^2}}{\sqrt{st}}\, (D_{X} W)^\dagger.
\eea

Defining
\be
F^T_0\equiv -4M_Pe^{\frac{1}{2}|X_i|^2}\sqrt{\frac{t^3}{s}}(D_T W)^\dagger, \label{ft0}
\ee
we can rewrite the F-term potential as
\bea
V_F&=&\frac{M^2_P|F^T_0|^2}{4t^2}+\frac{M^2_P|F^S|^2}{4s^2}+M^2_P|F^{X_i}|^2-\frac{3M^4_P|W|^2}{st} e^{|X_i|^2} \nonumber \\
&&+\frac{M^2_P|F^{Q'}|^2}{t}+\frac{M^2_P|F^{\varphi_i}|^2}{t}+\frac{M^2_P|F^{Q_i}|^2}{t} \nonumber \\
&&+\bigg[\frac{1}{t^2}Q_iQ^{'\dagger}F^{Q'}F^{Q_i^\dagger}+\frac{1}{t^2}Q_i\varphi^{\dagger}F^{\varphi}F^{Q_i^\dagger}
+\frac{1}{t^2}Q' \varphi^\dagger F^{\varphi} F^{Q'\dagger}+{\rm h.c.}\bigg]. \label{ftermpot1}
\eea

On the other hand, from eq.~(\ref{mostgeneral}), we obtain the scalar soft mass as follows,
we obtain the brane scalar soft mass as
\bea
m^2_i &=& \frac{1}{M^2_P}V_F+m^2_{3/2}-F^I F^{J\dagger}\partial_I\partial_{\bar J} \ln Z_i+\Big(-\frac{2}{t}+r_i\Big) g_R D_R
 \nonumber \\
&=&\frac{1}{M^2_P}\Big(\frac{2}{3}V_0-\frac{1}{3}sD^2_R\Big)-F^I F^{J\dagger}\partial_I\partial_{\bar J} \ln Y_i+\Big(-\frac{2}{t}+r_i\Big)g_R D_R.
\label{scalarmass}
\eea

\end{document}